Banner appropriate to article type will appear here in typeset article

# Floquet stability analysis of pulsatile particle-laden channel flow

**Ananthapadmanabhan Ramesh[1], Benoît Pier[2], and Parisa Mirbod[1]†**

[1]Department of Mechanical and Industrial Engineering, 842 W. Taylor Street, University of Illinois at Chicago, Chicago, IL 60607, USA

[2]Laboratoire de mécanique des fluides et d'acoustique, CNRS, École centrale de Lyon, Université de Lyon 1, INSA Lyon, 36 avenue Guy-de-Collongue, 69134 Écully, France

The linear stability of particle-laden pulsatile channel flow is investigated using Floquet analysis. The suspension consists of uniformly distributed spherical particles coupled to an incompressible Newtonian carrier fluid through Stokes drag and is modeled using a two-phase dusty-gas framework. The governing equations are linearized about a time-periodic base flow driven by a sinusoidally varying pressure gradient, and the effects of Reynolds number, Womersley number, pulsation amplitude, particle relaxation time, and particle mass fraction on temporal instability are systematically examined. In the steady limit, the influence of suspended particles depends strongly on their relaxation time: particles with very short relaxation times behave nearly as tracers and promote destabilization, whereas particles with finite relaxation times introduce interphase slip and drag-mediated damping, leading to stabilization. Under pulsatile forcing, the stability behavior becomes strongly frequency dependent. At low Womersley numbers, increasing pulsation amplitude destabilizes the flow, whereas at sufficiently large Womersley numbers it stabilizes the flow for all particle relaxation times considered. The results reveal a transition governed primarily by the penetration depth of oscillatory motion, with low-frequency forcing modulating the flow across the full channel and high-frequency forcing becoming increasingly confined to near-wall Stokes layers. Particle relaxation time and mass loading systematically shift this transition through interphase momentum exchange and drag-mediated attenuation. Evaluation of the transition boundary further shows that the corresponding values of $SWo^2$ remain small throughout the parameter space considered, indicating that particles remain strongly coupled to the carrier flow and that the observed transition is not associated with a resonance-like particle response. Instead, the stability behavior is governed by the coupled effects of oscillatory penetration and particle-fluid momentum exchange. These findings provide a unified framework for understanding instability in pulsatile particle-laden flows, with implications for physiological transport and periodically forced multiphase systems.

**Key words:** Hydrodynamic stability, Particle-fluid interactions, Pulsatile flow, Time periodic flows (Floquet analysis), Suspension dynamics

† Email address for correspondence: pmirbod@uic.edu

Abstract must not spill onto p.2

## 1. Introduction

Oscillatory shear flows occur when the driving pressure gradient or boundary motion varies periodically in time, introducing unsteady forcing that modifies the evolution of shear and hydrodynamic disturbances. Such flows are common in both biological and engineering settings. In physiological environments, pulsatile forcing governs cardiovascular hemodynamics, where periodic pumping of the heart generates oscillatory velocity fields that strongly influence transport, mixing, and flow stability (Ku 1997). Similar oscillatory transport also occurs in the motion of cerebrospinal fluid within the spinal canal (Loth *et al.* 2001). In engineering contexts, pulsatile flows are increasingly exploited to enhance heat and mass transfer, reduce membrane fouling, improve mixing efficiency, and control transport in microfluidic and chemical processing devices (Gillham *et al.* 2000; Rodrigues *et al.* 2015; Blythman *et al.* 2017; Dincau *et al.* 2020). In many such systems, periodic forcing substantially alters the transition behavior relative to steady flows by modifying the spatial penetration and temporal modulation of shear.

In practical applications, the carrier fluid frequently contains suspended particles, such as red blood cells in blood flow, aerosol particles in respiratory transport, or solid particles in industrial slurries (Melchionna 2011; Kleinstreuer & Zhang 2010; Roco & Shook 1983). The presence of dispersed particles modifies the flow through interphase momentum exchange and particle inertia, thereby altering disturbance amplification and transition thresholds. Understanding the stability of pulsatile particle-laden flows is therefore important for predicting transport behavior in a broad range of multiphase systems, including physiological transport, aerosol delivery, pulsed reactors, and suspension-based processing technologies.

The stability of pulsatile single-phase flows has been studied extensively. Early theoretical investigations examined the influence of time-periodic forcing on laminar Poiseuille flow. Grosch & Salwen (1968) analyzed the linear stability of plane Poiseuille flow subjected to a time-dependent pressure gradient and demonstrated that oscillatory forcing can substantially modify disturbance growth relative to steady conditions. Subsequent studies explored the effects of pulsation amplitude and frequency on stability characteristics. Hall (1975) investigated high-frequency modulation of Poiseuille flow, while Von Kerczek (1982) employed Floquet analysis to show that pulsation may either stabilize or destabilize channel flow depending on forcing conditions. Experimental investigations of oscillatory pipe flow similarly revealed complex frequency-dependent transition behaviour (Merkli & Thomann 1975; Hino *et al.* 1976; Stettler & Hussain 1986). More recent theoretical and numerical studies have further clarified the linear and nonlinear dynamics of pulsatile channel flows (Pier & Schmid 2017; Lebbal *et al.* 2022).

Parallel efforts have examined the stability of particle-laden shear flows using two-phase formulations, including dusty-gas models for dilute suspensions. Classical studies by Saffman (1962) and Michael (1964) demonstrated that suspended particles can significantly modify flow stability through inertial coupling between dispersed and carrier phases. Subsequent investigations showed that particle relaxation time and mass loading strongly influence disturbance growth and transition thresholds in channel flows (Rudyak & Isakov 1996; Rudyak *et al.* 1997; Klinkenberg *et al.* 2011, 2014). These studies established that suspended particles may either stabilize or destabilize shear flows depending on particle inertia, concentration, and interphase momentum exchange. More broadly, reviews of dispersed multiphase flows emphasize the central role of particle response and drag coupling in governing suspension dynamics across a wide range of flow regimes (Balachandar & Eaton 2010; Maxey 2017).

Despite these advances, the stability of pulsatile, particle-laden flows remains poorly understood. Most prior studies have focused on single-phase pulsatile flows or steady particle-laden channel flows, despite the prevalence of systems that combine periodic forcing and suspended particles. Examples include suspended cell-containing pulsatile blood flow, particle transport in oscillatory microfluidic devices, aerosol transport under cyclic forcing, and slurry transport in pulsed processing systems. In such flows, oscillatory forcing modifies the penetration of unsteady shear across the channel, while particle inertia and interphase drag alter disturbance dynamics through momentum exchange between the dispersed and carrier phases. How these competing effects jointly determine the transition between stabilization and destabilization remains largely unresolved. Existing work addressing this coupled problem is limited. For example, Chen & Chung (1995) showed that particles generally exert a destabilizing influence on oscillatory channel flow at a small Stokes number, with destabilization increasing with particle loading except at sufficiently high forcing frequencies.

Although particle response time naturally introduces an additional timescale into pulsatile multiphase flows, it remains unclear whether the transition between stabilization and destabilization is governed primarily by particle-response dynamics or by modifications to oscillatory shear penetration across the channel. In particular, the relative roles of oscillatory forcing, interphase drag coupling, and particle inertia in determining disturbance amplification remain unresolved. Clarifying these mechanisms is essential for understanding instability onset in pulsatile suspensions and for developing predictive descriptions of transition in periodically forced multiphase systems.

The present study investigates the linear stability of pulsatile particle-laden channel flow using a two-phase dusty-gas formulation. The governing equations are linearized about a time-periodic base flow generated by a sinusoidally varying pressure gradient, and Floquet theory is employed to characterize the evolution of infinitesimal disturbances under oscillatory forcing. The effects of pulsation amplitude, Womersley number, particle relaxation time, and particle mass fraction on temporal instability are systematically examined. Particular emphasis is placed on identifying how oscillatory forcing modifies instability by altering oscillatory penetration, and on how particle inertia and mass loading alter this behavior through interphase momentum coupling and drag-mediated damping. The central objective is to determine whether transition in pulsatile suspensions is controlled primarily by particle-response dynamics or by oscillatory penetration of the unsteady shear field.

We show that pulsation may either stabilize or destabilize the particle-laden channel flow, depending strongly on the Womersley number. Low-frequency forcing destabilizes the flow by modulating the shear across the channel, whereas sufficiently high-frequency forcing stabilizes the flow by confining oscillatory motion to near-wall regions. The particle relaxation time and mass loading systematically shift this transition by altering interphase coupling. More broadly, the analysis provides a predictive framework for understanding the mechanisms of instability in pulsatile multiphase shear flows relevant to physiological transport and periodically forced suspension systems. The remainder of the paper is organized as follows. Section 2 presents the problem formulation, including the governing equations, boundary conditions, and base state flow configuration. The linear stability framework and numerical methodology are described in Section 3. The stability results are presented and discussed in Section 4, with particular emphasis on the effects of the Womersley number, pulsation amplitude, particle mass fraction, and variations in the critical Reynolds number. Finally, Section 5 summarizes the main findings and provides directions for future work.

## 2. Problem formulation

We consider a pressure driven channel flow of a dilute suspension of small, rigid spherical particles confined between two parallel walls separated by a distance $2L$ as shown in figure 1. The flow is driven by a time periodic pressure gradient, which gives rise to a pulsatile base state. The carrier phase is modeled as an incompressible Newtonian fluid, while the dispersed phase consists of rigid spherical particles with diameters smaller than the smallest characteristic flow length scale (Klinkenberg *et al.* 2011). The suspension is described using a two phase dusty gas formulation, in which the fluid and particle phases are treated as interpenetrating continua coupled through interphase momentum exchange. This approach, originally introduced by Saffman (1962), is appropriate for dilute suspensions where particle–particle interactions, collisions, and finite volume effects can be neglected.

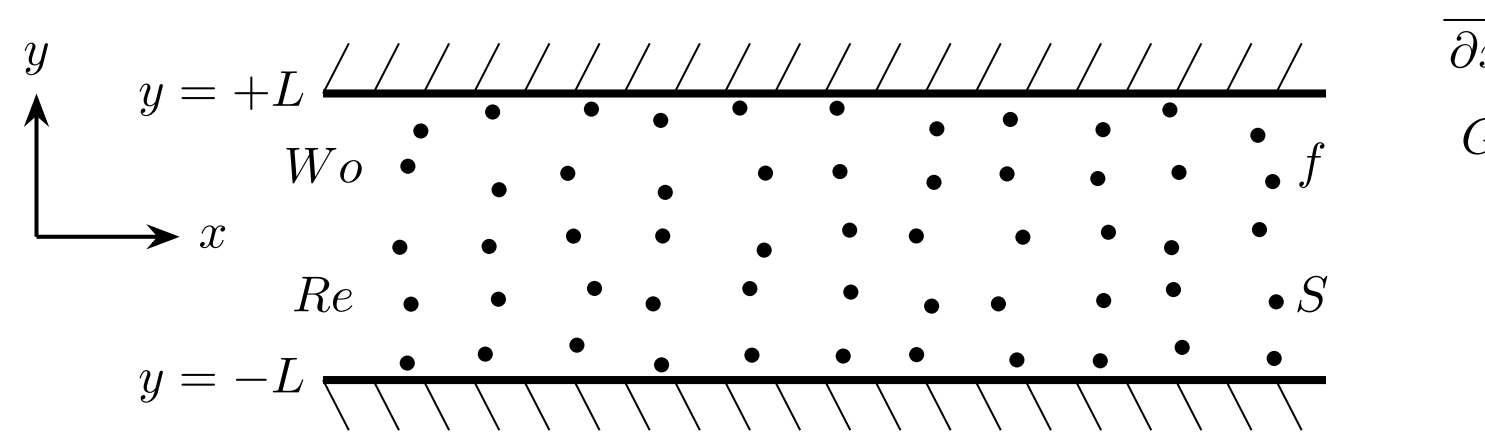

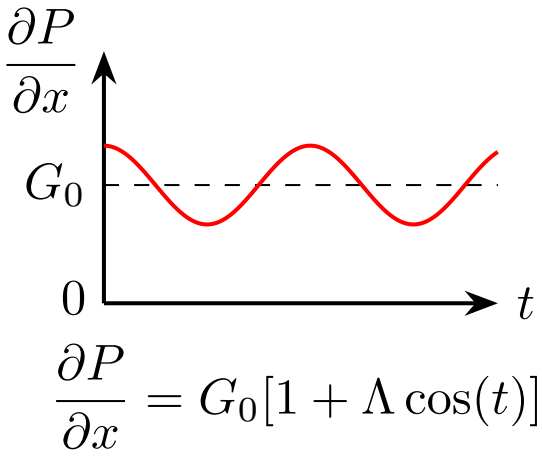

Figure 1: Schematic of pulsatile particle laden channel flow considered in the present study. A Newtonian carrier fluid driven by a sinusoidally varying pressure gradient transports uniformly distributed spherical particles characterized by mass fraction $f$ and relaxation time $S$ in a rigid channel of height $2L$. Other nondimensional parameters include $Re$ and $Wo$.

Particles in fluid systems are commonly classified as either heavy or light on the basis of their density relative to the carrier fluid. Heavy particles possess densities significantly greater than those of the fluid, whereas light particles have density ratios of order unity. For heavy particles, the fluid particle interaction is typically modeled considering only the Stokes drag force (Klinkenberg *et al.* 2013). In contrast, modeling the dynamics of light particles requires accounting for additional forces, including added mass, lift, buoyancy, fluid acceleration, and the Basset history force (Klinkenberg *et al.* 2014). We restricted our attention to single frequency forcing in order to isolate the fundamental interaction between oscillatory shear and particle relaxation. As a first step, we consider a dilute suspension and neglect particle–particle collisions and hydrodynamic interactions, while retaining two way momentum coupling between the phases. The formulation is therefore restricted to low particle volume fractions, for which such effects remain negligible. The continuum description adopted here is appropriate for laminar flow and infinitesimal disturbances, but is not intended for turbulent regimes in which particle clustering and finite concentration effects become important (Klinkenberg *et al.* 2014).

Three dimensional governing equations consist of the incompressible Navier Stokes equations for the fluid phase and a momentum balance for the particle phase, coupled through a drag force proportional to the slip velocity between the two phases. The dimensionalized governing equations for the fluid and particles are given as

$$\nabla \cdot \boldsymbol{u} = 0, \tag{2.1}$$

$$\rho\left(\frac{\partial \boldsymbol{u}}{\partial t} + \boldsymbol{u} \cdot \nabla \boldsymbol{u}\right) = -\nabla p + \mu \nabla^2 \boldsymbol{u} - n\boldsymbol{F}_{St}, \tag{2.2}$$

$$\frac{\partial n}{\partial t} + \nabla \cdot (n\boldsymbol{u_p}) = 0, \tag{2.3}$$

$$\frac{4\pi a^3}{3}\rho_p(\frac{\partial \boldsymbol{u_p}}{\partial t} + \boldsymbol{u_p}\cdot\nabla\boldsymbol{u_p}) = \boldsymbol{F}_{St}. \tag{2.4}$$

Here, $\boldsymbol{u}$, $\boldsymbol{u_p}$, $p$, $\rho$, $\rho_p$, $\mu$ and $n$ denote the velocity of the fluid, velocity of the particles, pressure, density of the fluid, density of the particles, dynamic viscosity and particle number density, respectively. The interaction between the fluid and the particles is governed by the Stokes drag force, expressed as $\boldsymbol{F}_{St} = 6\pi\mu a\,(\boldsymbol{u} - \boldsymbol{u_p})$, where $a$ is the radius of the spherical particles.

Equation 2.1 and 2.2 represent the continuity and momentum equations for the fluid phase, with the coupling between the fluid and the particle phases modeled through the term $n\boldsymbol{F}_{st}$. Similarly, equation 2.3 and 2.4 describe the transport and momentum of the particle phase. No slip boundary conditions are used for both the fluid and the particles at the top and bottom rigid walls.

$$\boldsymbol{u} = \boldsymbol{u_p} = 0; y = \pm L. \tag{2.5}$$

We nondimensionalize length by the channel half width (L), velocity by the centreline velocity ($U_{\max}$) of steady plane Poiseuille flow, time by the forcing time scale ($1/\omega$), where $\omega$ is the angular frequency of the imposed pulsation, particle number density by ($N_0$), and pressure by ($\rho U_{\max}^2$). These scaling factors allow the governing equations to be expressed in a dimensionless form, facilitating analysis and comparison between different flow conditions. Equations 2.6 - 2.13 present the non-dimensionalized continuity and momentum formulations for the fluid and particle phases driven by a pulsatile pressure gradient, in which several new dimensionless parameters are introduced. The flow field is controlled by the Reynolds number ($Re$), defined as $\rho U_{max}L/\mu$ and the Womersley number ($Wo$) defined as $Wo = L\sqrt{\frac{\omega}{\nu}}$. The parameter $f$ (mass fraction) is expressed as $4\pi a^3\rho_p N_0/3\rho$, where $a$ is the radius of the particle. The dimensionless particle relaxation time is defined as $S$= $2\rho_p a^2/9\rho L^2$. The Womersley number characterizes the relative importance of unsteady inertial effects and viscous diffusion, while the parameter $S$ represents the particle response time relative to the characteristic time scale of the flow.

The dimensionless governing equations for the fluid and particle phases are given by the equations below, where the coupling terms explicitly account for momentum exchange between the phases.

$$\frac{\partial u}{\partial x} + \frac{\partial v}{\partial y} + \frac{\partial w}{\partial z} = 0 \tag{2.6}$$

$$\frac{Wo^2}{Re}\frac{\partial u}{\partial t} + u\frac{\partial u}{\partial x} + v\frac{\partial u}{\partial y} + w\frac{\partial u}{\partial z} = -\frac{\partial P}{\partial x} + \frac{1}{Re}\left(\frac{\partial^2 u}{\partial x^2} + \frac{\partial^2 u}{\partial y^2} + \frac{\partial^2 u}{\partial z^2}\right) - \frac{fn}{SRe}(u - u_p) \tag{2.7}$$

$$\frac{Wo^2}{Re}\frac{\partial v}{\partial t} + u\frac{\partial v}{\partial x} + v\frac{\partial v}{\partial y} + w\frac{\partial v}{\partial z} = -\frac{\partial P}{\partial y} + \frac{1}{Re}\left(\frac{\partial^2 v}{\partial x^2} + \frac{\partial^2 v}{\partial y^2} + \frac{\partial^2 v}{\partial z^2}\right) - \frac{fn}{SRe}(v - v_p) \tag{2.8}$$

$$\frac{Wo^2}{Re}\frac{\partial w}{\partial t} + u\frac{\partial w}{\partial x} + v\frac{\partial w}{\partial y} + w\frac{\partial w}{\partial z} = -\frac{\partial P}{\partial z} + \frac{1}{Re}\left(\frac{\partial^2 w}{\partial x^2} + \frac{\partial^2 w}{\partial y^2} + \frac{\partial^2 w}{\partial z^2}\right) - \frac{fn}{SRe}(w - w_p) \tag{2.9}$$

$$\frac{Wo^2}{Re}\frac{\partial n}{\partial t} + u_p\frac{\partial n}{\partial x} + v_p\frac{\partial n}{\partial y} + w_p\frac{\partial n}{\partial z} + n\left(\frac{\partial u_p}{\partial x} + \frac{\partial v_p}{\partial y} + \frac{\partial w_p}{\partial z}\right) = 0 \tag{2.10}$$

$$\frac{Wo^2}{Re}\frac{\partial u_p}{\partial t} + u_p\frac{\partial u_p}{\partial x} + v_p\frac{\partial u_p}{\partial y} + w_p\frac{\partial u_p}{\partial z} = \frac{1}{SRe}(u - u_p) \tag{2.11}$$

$$\frac{Wo^2}{Re}\frac{\partial v_p}{\partial t} + u_p\frac{\partial v_p}{\partial x} + v_p\frac{\partial v_p}{\partial y} + w_p\frac{\partial v_p}{\partial z} = \frac{1}{SRe}(v - v_p) \tag{2.12}$$

$$\frac{Wo^2}{Re}\frac{\partial w_p}{\partial t} + u_p\frac{\partial w_p}{\partial x} + v_p\frac{\partial w_p}{\partial y} + w_p\frac{\partial w_p}{\partial z} = \frac{1}{SRe}(w - w_p) \tag{2.13}$$

In particular, the interphase drag introduces a relaxation mechanism that tends to reduce the slip velocity between the fluid and particle phases over a characteristic time scale proportional to $S$. The dimensionless boundary conditions are given by

$$\begin{aligned} u = v = w = 0 \quad \text{at } y = \pm 1 \\ u_p = v_p = w_p = 0 \quad \text{at } y = \pm 1. \end{aligned} \tag{2.14}$$

The flow is driven by a spatially uniform, time periodic pressure gradient of the form

$$\frac{\partial P}{\partial x} = G_0(1 + \Lambda cos(t)) = G_0\left(1 + \Lambda\frac{\exp(it) + \exp(-it)}{2}\right); \tag{2.15}$$

Here, $G_0$ is the mean pressure gradient driving the flow, and $\Lambda$ denotes the dimensionless amplitude of the oscillatory forcing relative to the steady component. This forcing produces a base flow that is periodic in time with period $2\pi$. The corresponding volumetric flow rate is also periodic and can be expressed as

$$Q(t) = Q_0(1 + \delta cos(t)); \tag{2.16}$$

where $Q_0$ denotes the mean (steady) volumetric flow rate, while $\delta$ represents the dimensionless amplitude of the oscillatory component relative to the mean flow rate. In the present study, attention is restricted to a single frequency forcing in order to isolate the fundamental mechanisms governing pulsatile particle laden flows. This simplification allows a systematic investigation of the interaction between oscillatory forcing and particle inertia without introducing additional complexity associated with multi frequency excitation. Physically, the problem is controlled by the ratio of the oscillation time scale to the particle relaxation time, which determines the phase relationship between fluid and particle motion and hence the strength of interphase momentum exchange. The relative magnitude of these time scales plays a central role in determining the evolution of disturbances, as it controls the phase relationship between fluid and particle motion and therefore the effectiveness of momentum coupling between the phases.

### 2.1. *Base state velocity*

The base state velocity profiles for the fluid and particle phases are obtained by solving the governing equations under the assumption of spatially parallel flow. Both phases exhibit streamwise velocities that depend only on the wall normal coordinate $y$ and time $t$. The fluid base flow can be decomposed into a steady and an unsteady component,

$$U(y, t) = \bar{U}(y) + U_1(y, t) \tag{2.17}$$

where the steady contribution corresponds to the classical plane Poiseuille flow,

$$\bar{U}(y) = 1 - y^2 \tag{2.18}$$

and the unsteady component arises from the oscillatory forcing. The unsteady velocity field can be expressed in terms of harmonic contributions,

$$U_1(y,t) = \frac{\Lambda}{A^2}\left(1 - \frac{\cosh Ay}{\cosh A}\right)e^{it} + \frac{\Lambda}{A^{*2}}\left(1 - \frac{\cosh A^* y}{\cosh A^*}\right)e^{-it}$$
$$= q_1(y)e^{it} + q_{-1}(y)e^{-it} \tag{2.19}$$

with $A^*$ denoting the complex conjugate of $A$. The parameter $A$ depends on the Womersley number and the particle coupling and governs the depth of penetration of the oscillatory motion in the flow. It is defined as:

$$A = \left(\frac{fn}{S} - \frac{fn}{S(1 + iSWo^2)} + iWo^2\right)^{0.5}. \tag{2.20}$$

In the present study, a uniform particle distribution is considered in the base state, implying a constant particle number density. Upon nondimensionalization with reference density $N_0$, the base state value is given by $n = 1$ (Boronin & Osiptsov 2014; Kumar & Govindarajan 2024). Similarly, the base state velocity of the particle phase can be written as

$$U_p(y,t) = \bar{U}(y) + U_{p1}(y,t) \tag{2.21}$$

with the unsteady component given by

$$U_{p1}(y,t) = \frac{\Lambda}{A^2(1 + iSWo^2)}\left(1 - \frac{\cosh Ay}{\cosh A}\right)e^{it} + \frac{\Lambda}{A^{*2}(1 - iSWo^2)}\left(1 - \frac{\cosh A^* y}{\cosh A^*}\right)e^{-it}$$
$$= q_2(y)e^{it} + q_{-2}(y)e^{-it} \tag{2.22}$$

The particle response is governed by the factor $(1 \pm iSWo^2)^{-1}$, which reflects the influence of particle relaxation time relative to the imposed oscillatory forcing. As $SWo^2$ increases, finite particle inertia may introduce attenuation and particle-fluid phase differences. In the present parameter range, however, $SWo^2$ remains small, indicating strong fluid-particle coupling and suggesting that large particle–fluid phase lag is not expected. Consequently, particle inertia primarily modifies the instability behaviour through interphase momentum exchange and drag-mediated damping.

The influence of pulsation can be characterized through the amplitudes of both the imposed pressure gradient and the resulting flow rate. While the pulsating pressure gradient serves as the driving mechanism, it is the velocity field, or equivalently, the flow rate, that directly defines the base state in the stability equations. The flow rate provides a direct measure of the extent to which the velocity field oscillates in time. We therefore use the flow rate amplitude $\delta$ as the principal measure of pulsation intensity in the stability analysis. The results presented below are expressed entirely in dimensionless form. Unless stated otherwise, velocities are normalized by $U_{\max}$, time by $1/\omega$, and perturbation amplitudes in temporal traces are normalized by the initial perturbation kinetic energy. These conventions are used consistently in the figures and captions.

Although the amplitude of the pressure gradient $\Lambda$ can also be used to describe the pulsation, the two quantities are related through the equation 2.23, allowing for conversion when necessary.

$$\frac{\delta}{\Lambda} = \frac{3}{2}\left(\frac{e^{it}}{A^2}\left(1 - \frac{tanhA}{A}\right) + \frac{e^{-it}}{A^{*2}}\left(1 - \frac{tanhA^*}{A^*}\right)\right) \tag{2.23}$$

Figures 2 illustrate the base state velocity profiles of the fluid and particle phases over one forcing cycle for representative values of $Wo$ = 5 and the pulsation amplitude. At this relatively low Womersley number, viscous diffusion dominates over unsteady inertial effects, allowing the oscillatory component of the flow to penetrate across the entire channel. As a result, both the fluid and the particle velocity profiles exhibit significant temporal modulation in the entire domain, with noticeable differences between the phases ($t = \pi / 2$) and ($t=3 \pi / 2$). For the fluid phase, the velocity profile remains parabolic with highest peak velocity for $t= \pi/2$ and lowest at $t= 3\pi/2$ as shown in figure 2a. An increase in pulsating amplitude enhances the magnitude of base state velocity, without altering the phase dependent variation as shown in figure 2b. The particle phase follows a similar trend, but with a reduced amplitude and a slight phase lag relative to the fluid. This difference reflects the finite response time of the particles and highlights the role of particle inertia in modulating the base flow.

The analysis is further extended to higher $Wo$=25, as shown in figure 3. At this higher Womersley number, unsteady inertial effects dominate over viscous diffusion, leading to the confinement of the oscillatory component within thin near wall Stokes layers. For the fluid phase, the velocity profile remains close to parabolic, with a maximum at $t=\pi/2$ and a minimum at $t=3\pi/2$. However, the difference between these profiles is smaller than that observed at low $Wo$, indicating reduced temporal modulation in the bulk of the flow. The oscillatory effects are primarily localized near the walls, where steeper velocity gradients develop as a result of the thin Stokes layer. For the particle phase, the velocity profiles closely follow the fluid motion, but exhibit slightly reduced amplitude, reflecting the finite particle response time.

## 3. Linear stability analysis

To investigate the stability of the particle laden pulsatile channel flow, small amplitude perturbations are superimposed on the time periodic base state. Substitution of perturbed variables into the governing equations and subsequent linearization yield a system of linear equations that describe the evolution of infinitesimal disturbances (denoted $'$). The flow variables are decomposed into base and perturbation components as follows:

Fluid phase:

$$\begin{aligned} &u(x,y,t) = U(y,t) + u'(x,y,t), \quad v(x,y,t) = v'(x,y,t), \\ &p(x,y,t) = P(x,t) + p'(x,y,t) \end{aligned} \tag{3.1}$$

Particle phase:

$$\begin{aligned} &u_p(x,y,t) = U_p(y,t) + u'_p(x,y,t), \quad v_p(x,y,t) = v'_p(x,y,t), \\ &n(x,y,t) = N_0 + n'(x,y,t) \end{aligned} \tag{3.2}$$

where primed quantities denote small perturbations. After substituting the perturbed dependent variables into the governing equations of fluid flow, the following linearized equations are obtained as,

$$\frac{\partial u'}{\partial x} + \frac{\partial v'}{\partial y} + \frac{\partial w'}{\partial z} = 0 \tag{3.3}$$

$$\frac{Wo^2}{Re}\frac{\partial u'}{\partial t} + U\frac{\partial u'}{\partial x} + \frac{dU}{dy}v' = -\frac{\partial p'}{\partial x} + \frac{1}{Re}\left(\frac{\partial^2 u'}{\partial x^2} + \frac{\partial^2 u'}{\partial y^2} + \frac{\partial^2 u'}{\partial z^2}\right) - \frac{fN_0}{SRe}(u' - u'_p) - \frac{f}{SRe}(U - U_p)n' \tag{3.4}$$

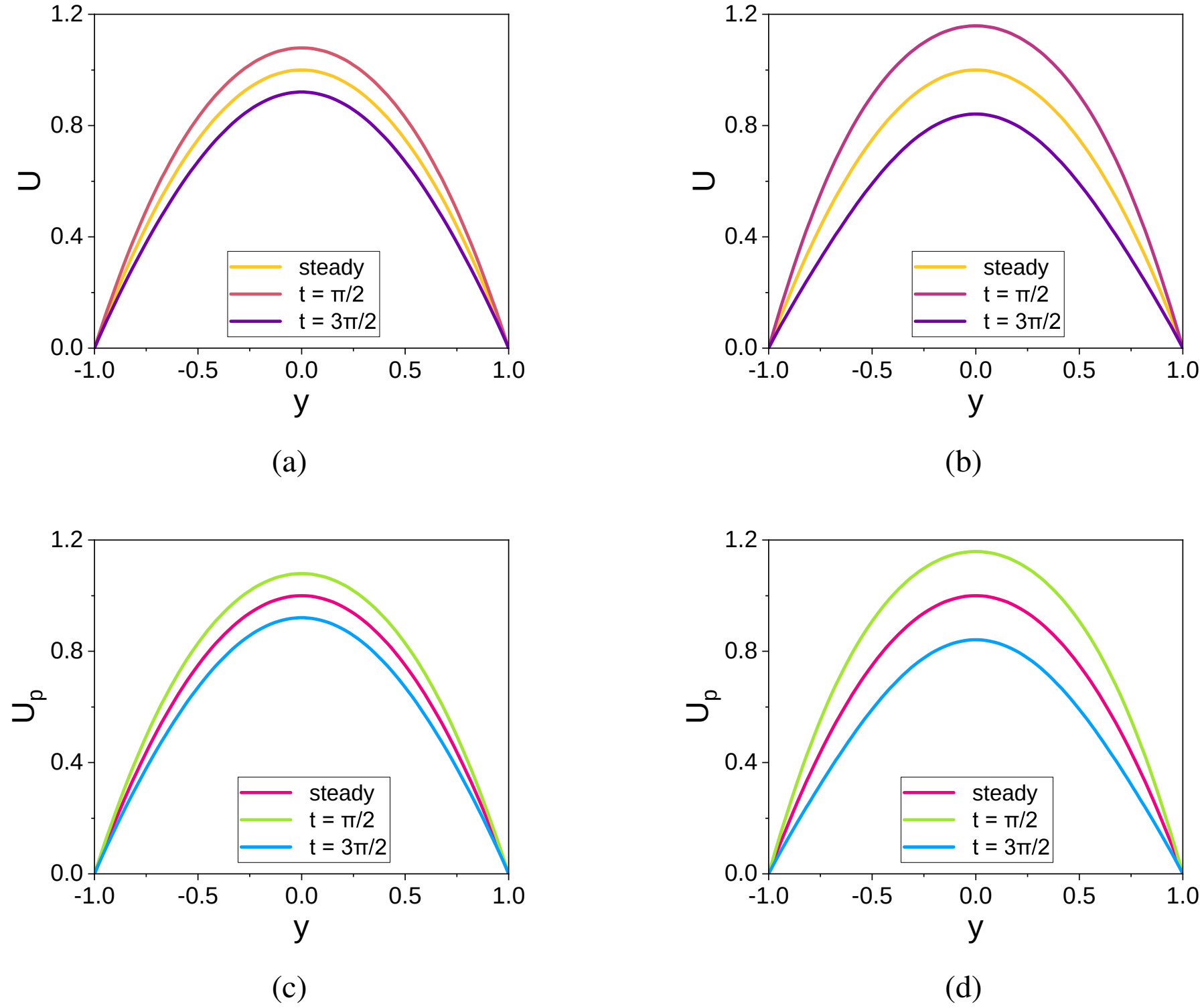

Figure 2: Pulsating base state velocity profiles of fluid and particles presented in terms of the dimensionless velocity with $Wo$=5, $f$=0.1, $S = 5 \times 10^{-5}$. (a,b) Fluid velocity with $\delta$=0.1 and $\delta$=0.2; (c,d) Particle velocity with $\delta$=0.1 and $\delta$=0.2. The oscillatory component penetrates across most of the channel, producing temporally modulated shear throughout the bulk flow.

$$\frac{Wo^2}{Re}\frac{\partial v'}{\partial t} + U\frac{\partial v'}{\partial x} = -\frac{\partial p'}{\partial y} + \frac{1}{Re}\left(\frac{\partial^2 v'}{\partial x^2} + \frac{\partial^2 v'}{\partial y^2} + \frac{\partial^2 v'}{\partial z^2}\right) - \frac{fN_0}{SRe}(v' - v'_p) \tag{3.5}$$

$$\frac{Wo^2}{Re}\frac{\partial w'}{\partial t} + U\frac{\partial w'}{\partial x} = -\frac{\partial p'}{\partial z} + \frac{1}{Re}\left(\frac{\partial^2 w'}{\partial x^2} + \frac{\partial^2 w'}{\partial y^2} + \frac{\partial^2 w'}{\partial z^2}\right) - \frac{fN_0}{SRe}(w' - w'_p) \tag{3.6}$$

The perturbed equations for the particle phase are derived by linearizing the governing equations for the particle dynamics around the base state. The corresponding dimensionless continuity and momentum equations for the particle phase are given as

$$\frac{Wo^2}{Re}\frac{\partial n'}{\partial t} + U_p\frac{\partial n'}{\partial x} + \frac{\partial u'_p}{\partial x} + \frac{\partial v'_p}{\partial y} + \frac{\partial w'_p}{\partial z} = 0, \tag{3.7}$$

$$\frac{Wo^2}{Re}\frac{\partial u'_p}{\partial t} + U_p\frac{\partial u'_p}{\partial x} + \frac{\partial U_p}{\partial y}v'_p = \frac{1}{SRe}(u' - u'_p) \tag{3.8}$$

$$\frac{Wo^2}{Re}\frac{\partial v'_p}{\partial t} + U_p\frac{\partial v'_p}{\partial x} = \frac{1}{SRe}(v' - v'_p) \tag{3.9}$$

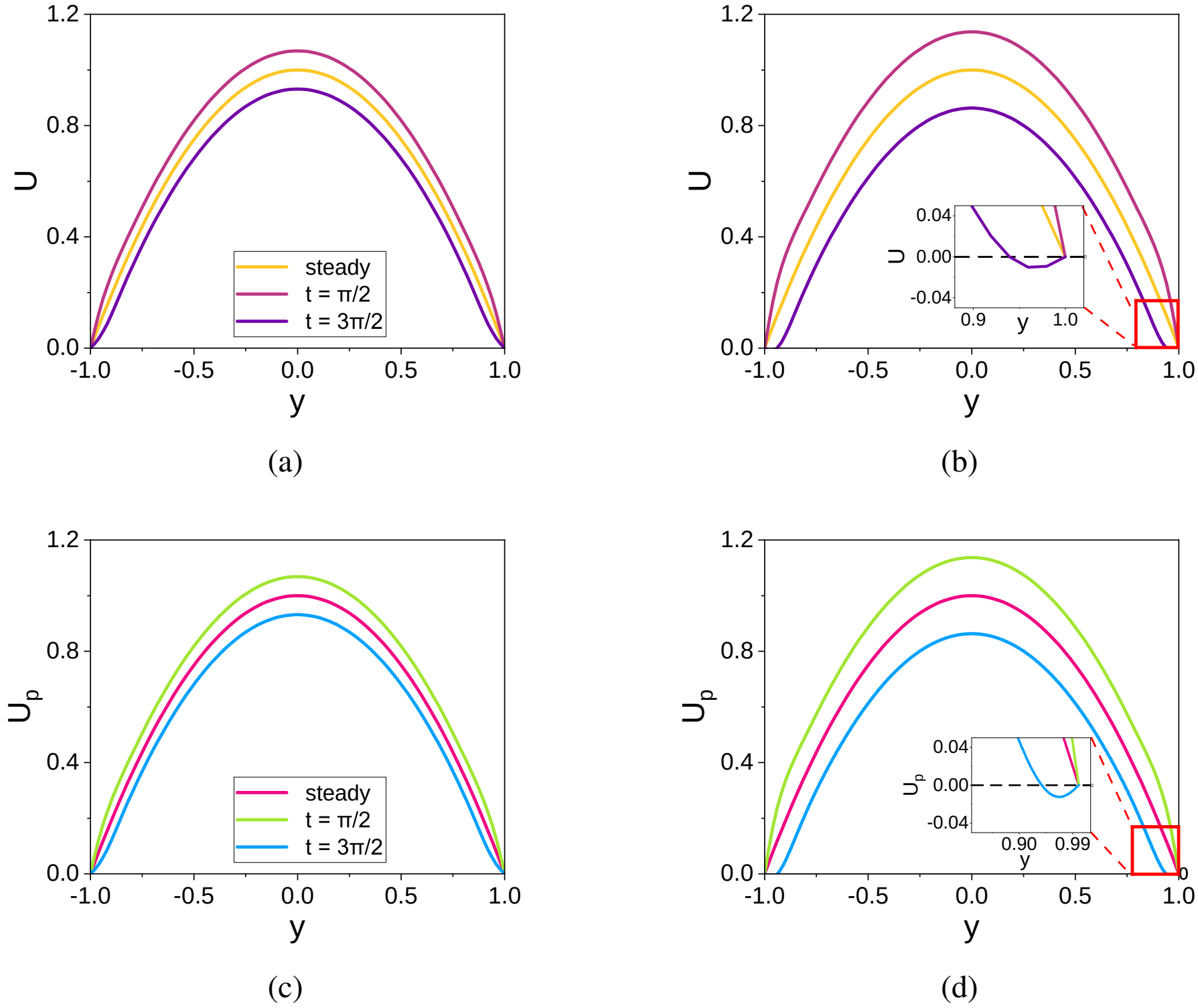

Figure 3: Pulsating base state velocity profiles of fluid and particles shown in terms of the dimensionless velocity for $Wo$=25, $f$=0.1, $S = 5 \times 10^{-5}$. (a,b) Fluid velocity with $\delta$=0.1 and $\delta$=0.2; (c,d) Particle velocity with $\delta$=0.1 and $\delta$=0.2. At larger $Wo$, oscillatory motion becomes confined to thin near-wall Stokes layers, while the channel core experiences weak temporal modulation.

$$\frac{Wo^2}{Re}\frac{\partial w'_p}{\partial t} + U_p\frac{\partial w'_p}{\partial x} = \frac{1}{SRe}(w' - w'_p) \tag{3.10}$$

The corresponding boundary condition for fluid and particle phase after perturbation can be expressed as

$$\begin{aligned} u' = v' = w' = 0 \quad &\text{at } y = \pm 1 \\ u'_p = v'_p = w'_p = 0 \quad &\text{at } y = \pm 1 \end{aligned} \tag{3.11}$$

Considering three dimensional perturbation of the form $\xi'(x, y, z, t) = \hat{\xi}(y, t)e^{i(\alpha x+\beta z)}$, where $\hat{\xi}(y) = (\hat{u}, \hat{v}, \hat{w}, \hat{u}_p, \hat{v}_p, \hat{w}_p, \hat{n}, \hat{p})$ denotes the amplitude of perturbations for the fluid and particle phases. Using Squire's theorem (see appendix A), it can be shown that two dimensional disturbances become unstable at Reynolds numbers that are less than or equal to those required for fully three dimensional disturbances. Therefore, without loss of generality, the modal linear stability analysis of particle laden pulsatile channel flow can be restricted to two dimensional perturbations, corresponding to $\alpha$=$k$ and $\beta$=0. A rigorous extension of Squire's theorem to time periodic Floquet systems is non trivial. However, under the present assumptions, the transformed disturbance equations retain the structure of a two dimensional problem, and the least stable modes are therefore expected to remain two dimensional. The following analysis is restricted accordingly.

Restricting attention to two dimensional disturbances, we introduce the stream function $\psi$, defined by $u' = \frac{\partial \psi}{\partial y}$ and $v' = -\frac{\partial \psi}{\partial x}$, which automatically satisfies continuity. This formulation automatically satisfies the continuity equation and allows the governing equations to be reduced to a system involving $\psi$, particle velocity perturbations, and particle concentration. Hence, the governing perturbation equations 3.3 to 3.9 can be further reduced, as shown below.

$$\frac{Wo^2}{Re}\frac{\partial}{\partial t}\left(\frac{\partial^2 \psi}{\partial x^2}+\frac{\partial^2 \psi}{\partial y^2}\right)+U\frac{\partial^3 \psi}{\partial x \partial y^2}-\frac{d^2 U}{dy^2}\frac{\partial \psi}{\partial x}+U\frac{\partial^3 \psi}{\partial x^3}=\frac{1}{Re}\left(\frac{\partial^4 \psi}{\partial x^4}+2\frac{\partial^4 \psi}{\partial x^2 y^2}+\frac{\partial^4 \psi}{\partial y^4}\right)$$
$$-\frac{fN_0}{SRe}\left(\frac{\partial^2 \psi}{\partial y^2}-\frac{\partial u'_p}{\partial y}\right)-\frac{\partial n'}{\partial y}\frac{f}{SRe}(U-U_p)-\frac{fn'}{SRe}\left(\frac{\partial U}{\partial y}-\frac{\partial U_p}{\partial y}\right)$$
$$-\frac{fN_0}{SRe}\left(\frac{\partial^2 \psi}{\partial x^2}+\frac{\partial v'_p}{\partial x}\right) \quad (3.12)$$

$$\frac{Wo^2}{Re}\frac{\partial n'}{\partial t}+U_p\frac{\partial n'}{\partial x}+N_0\frac{\partial u'_p}{\partial x}+N_0\frac{\partial v'_p}{\partial y}=0 \quad (3.13)$$

$$\frac{Wo^2}{Re}\frac{\partial u'_p}{\partial t}+U_p\frac{\partial u'_p}{\partial x}+\frac{\partial U_p}{\partial y}v'_p=\frac{1}{SRe}\left(\frac{\partial \psi}{\partial y}-u'_p\right) \quad (3.14)$$

$$\frac{Wo^2}{Re}\frac{\partial v'_p}{\partial t}+U_p\frac{\partial v'_p}{\partial x}=-\frac{1}{SRe}\left(\frac{\partial \psi}{\partial x}+v'_p\right) \quad (3.15)$$

The coupling between fluid and particle perturbations is mediated by the drag terms proportional to $(u' - u'_p)$ and $(v' - v'_p)$, which act to relax the slip velocity between the phases over a characteristic time scale proportional to $S$. This coupling introduces an additional mechanism influencing disturbance evolution: while the fluid responds directly to the oscillatory base flow, the particle phase responds with a delay governed by the relaxation time. As a result, perturbations in the particle phase may either damp or amplify fluid disturbances depending on the relative slip between the two phases.

Under the assumption that the particle relaxation time is sufficiently small, the base state velocities of the fluid and particle phases become nearly identical ($U \approx U_p$) (Chen & Chung 1995). In this limit, the coupling terms involving $n'$ disappear from the governing equations, and the particle concentration decouples from the momentum equations. Consequently, the perturbation system reduces to a coupled set of equations for the fluid and particle velocities, independent of the particle continuity equation. This simplification significantly reduces the complexity of the stability analysis while retaining the essential effects of particle inertia.

### 3.1. *Floquet Analysis*

Because the base flow is periodic in time, classical normal mode analysis is not directly applicable. Instead, Floquet theory is employed to characterize the stability of disturbances in time periodic systems. In this framework, perturbations are represented as exponential growth modulated by time periodic functions, allowing the determination of stability through Floquet exponents. Accordingly, the perturbation variables are expanded in Fourier modes in time and normal modes in the streamwise direction. The unknowns are expressed as a product of a spatial wave, an exponential growth term, and a sum of temporal harmonics. Specifically, each perturbation quantity is written as a superposition of modes of the form,

$$\begin{aligned} \psi' &= e^{ikx} e^{\sigma t} \sum_{m=-\infty}^{+\infty} \hat{\psi}_m e^{imt}; \\ u'_p &= e^{ikx} e^{\sigma t} \sum_{m=-\infty}^{+\infty} \hat{u}_{p,m} e^{imt}; \\ v'_p &= e^{ikx} e^{\sigma t} \sum_{m=-\infty}^{+\infty} \hat{v}_{p,m} e^{imt}; \end{aligned} \tag{3.16}$$

where $\sigma$ is the complex growth rate, whose real value defines temporal instability(+ve)/ stability(-ve) of the system and $k$ is dimensionless perturbation wavenumber.

We substitute the Floquet expanded perturbation variables given in equation 3.16 into the governing perturbation equations (equations 3.12-3.15). This procedure transforms the time periodic system into a set of coupled equations for the harmonic modes. Accordingly, the governing equation for the streamfunction amplitude $\psi$ can be written as

$$\begin{aligned} &\frac{imWo^2}{Re}\left(\frac{\partial^2}{\partial y^2} - k^2\right)\hat{\psi}_m + \bar{U}ik\left(\frac{\partial^2}{\partial y^2} - k^2\right)\hat{\psi}_m - ik\frac{d^2\bar{U}}{dy^2}\hat{\psi}_m - \frac{1}{Re}\left(\frac{\partial^4}{\partial x^4} - 2k^2\frac{\partial^2}{\partial y^2} + k^4\right)\hat{\psi}_m \\ &+ \frac{fN_0}{SRe}\left(\frac{\partial^2}{\partial y^2} - k^2\right)\hat{\psi}_m + ik\left(q_1\left(\frac{\partial^2}{\partial y^2} - k^2\right) - \frac{d^2q_1}{dy^2}\right)\hat{\psi}_{m-1} + \\ &ik\left(q_{-1}\left(\frac{\partial^2}{\partial y^2} - k^2\right) - \frac{d^2q_{-1}}{dy^2}\right)\hat{\psi}_{m+1} - \left(\frac{fN_0}{SRe}\frac{\partial}{\partial y}\right)\hat{u}_{p,m} + \frac{fN_0}{SRe}ik\hat{v}_{p,m} \\ &= -\sigma\frac{Wo^2}{Re}\left(\frac{\partial^2}{\partial y^2} - k^2\right)\hat{\psi}_m \end{aligned} \tag{3.17}$$

Similarly, the governing equations for the particle velocity components $u_{p,m}$ and $v_{p,m}$ are obtained as

$$\begin{aligned} &\frac{-1}{SRe}\frac{\partial\hat{\psi}_m}{\partial y} + \left(\frac{Wo^2 im}{Re} + \bar{U}ik + \frac{1}{SRe}\right)\hat{u}_{p,m} + q_2 ik\hat{u}_{p,m-1} + q_{-2}ik\hat{u}_{p,m+1} \\ &+ \frac{d\bar{U}}{dy}\hat{v}_{p,m} + \frac{dq_2}{dy}\hat{v}_{p,m-1} + \frac{dq_{-2}}{dy}\hat{v}_{p,m+1} = -\sigma\frac{Wo^2}{Re}\hat{u}_{p,m} \end{aligned} \tag{3.18}$$

$$ik\frac{1}{SRe}\hat{\psi}_m + \left(\frac{Wo^2 im}{Re} + \bar{U}ik + \frac{1}{SRe}\right)\hat{v}_{p,m} + q_2 ik\hat{v}_{p,m-1} + q_{-2}ik\hat{v}_{p,m+1} = -\sigma\frac{Wo^2}{Re}\hat{v}_{p,m} \tag{3.19}$$

These equations characterize how the particle phase dynamically adjusts to the motion of the carrier fluid, where the strength of interaction is controlled by the particle parameters $f$ and $S$. For the case of a uniform particle distribution across the channel, the number density becomes constant, i.e. $N_0$ =1. Under this assumption, the governing equations simplify significantly, as the spatial variation of particle concentration no longer contributes to the coupling terms. This allows us to isolate the effect of particle inertia and drag on the flow stability. The boundary conditions may be recast in terms of the streamfunction formulation. Using the relations between velocity components and streamfunction, the no slip and no penetration

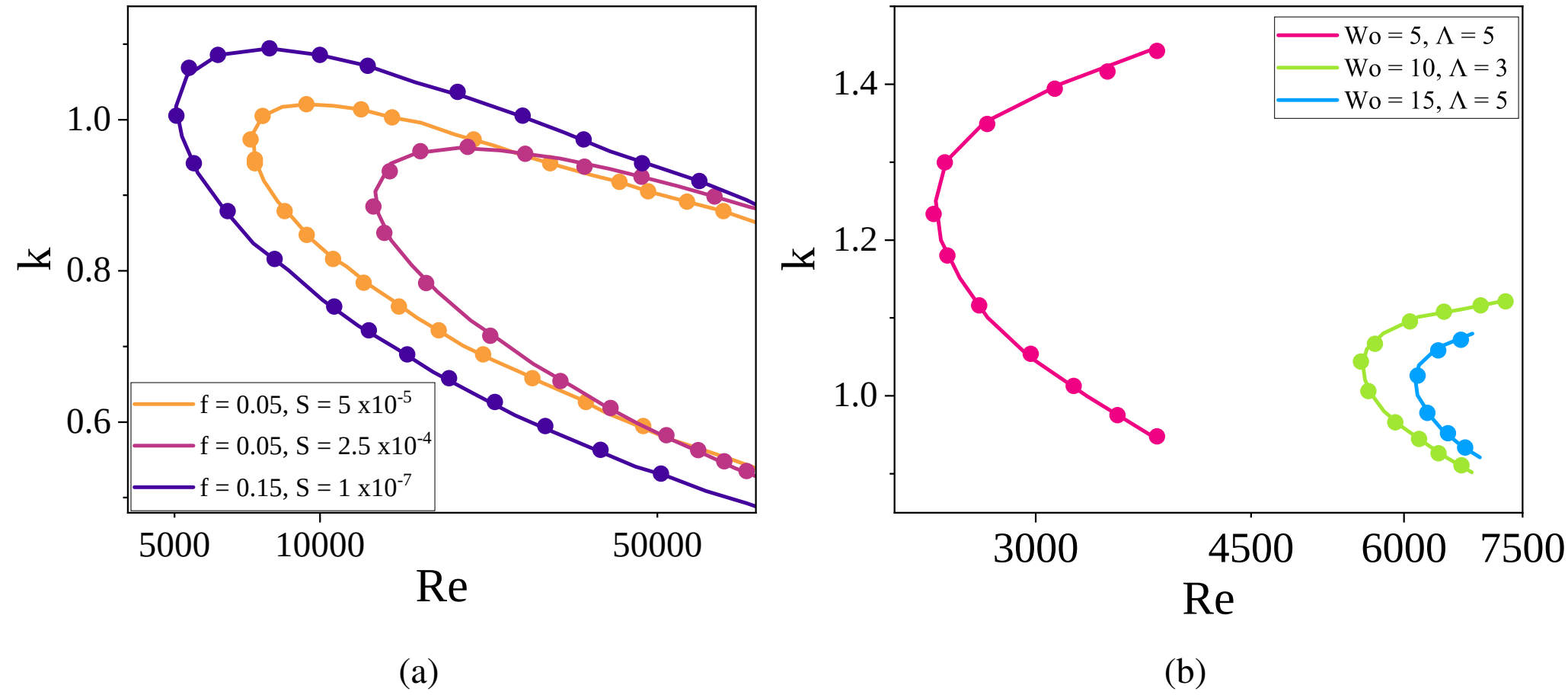

(a) (b)

Figure 4: Validation of the model developed in the form of neutral stability curves. (a) Comparison of the current work with Klinkenberg *et al.* (2011) for different $f$ and $S$; (b) Comparison of the current work with Tsigklifis & Lucey (2017) for different Womersely numbers and $\Lambda$. Symbols correspond to present computations, and solid lines are from the reference papers.

conditions at the walls yield

$$\hat{\psi}_m = D\hat{\psi}_m = 0 \quad \text{at } y = \pm 1, \tag{3.20}$$

where $D = \frac{\partial}{\partial y}$. The particle phase boundary conditions remain unchanged and are given by

$$u = u_p = v = v_p = 0 \quad \text{at } y = \pm 1. \tag{3.21}$$

### 3.2. *Numerical method*

The linear stability problem derived in the previous section leads to a generalized eigenvalue problem of the form

$$AX = \sigma BX,$$

where $\sigma$ is the complex growth rate and $X$ contains the amplitudes of the Fourier modes associated with the fluid and particle perturbations. In the wall normal direction, the governing equations are discretized using a Chebyshev spectral collocation method. The perturbation fields are approximated using truncated expansions of Chebyshev polynomials, evaluated at the Chebyshev Gauss Lobatto nodes $y(j) = cos\left(\frac{\pi j}{J}\right)$, where j = 0,1,,,,J, which cluster near the channel walls and allow accurate resolution of boundary layers and strong velocity gradients. The wall normal differential operators are thus represented by Chebyshev differentiation matrices acting on the discrete solution vectors. In the temporal direction, the Floquet expansion is truncated to a finite number of harmonics. The infinite Fourier series is approximated using the $2M + 1$ modes, corresponding to the harmonic indices $m = -M, \ldots, M$. This truncation results in a coupled system of equations for all retained modes, reflecting the interaction between neighboring harmonics induced by the time periodic base flow.

The Chebyshev discretization in the wall normal direction combined with the truncated Floquet expansion in time yields a block structured generalized eigenvalue problem, which is solved using standard MATLAB eigensolvers. Temporal stability is determined by the eigenvalue with the largest real part. A systematic convergence study was conducted to ensure numerical accuracy. The sensitivity of the leading eigenvalue to both spatial resolution $J$ and temporal resolution $M$ was examined. For a representative case with Womersley number $Wo = 15$ and flow rate amplitude $\delta = 0.1$, the growth rate of the leading mode was calculated for increasing values of $J$ and $M$. The results, summarized in Table 1, demonstrate rapid convergence of the eigenvalues with increasing resolution. In particular, the solutions obtained with $J = 70$ and $J = 100$ agree at six decimal places, indicating spectral convergence in the wall normal direction. The convergence with respect to the number of Fourier modes was similarly verified. Based on this analysis, all subsequent computations were performed using $70 \leq J \leq 100$ and $250 \leq M \leq 300$, which ensures accurate resolution of both spatial gradients and temporal harmonic interactions.

| $J$ | $M$ | $\sigma_r$ | $\sigma_i$ |
|---|---|---|---|
| 50 | 50 | -0.005353356 | 0.2338302 |
| 50 | 100 | -0.005353356 | -0.766169 |
| 70 | 150 | -0.005484924 | -0.7661589 |
| 70 | 250 | -0.005484924 | -0.7661589 |
| 100 | 350 | -0.005484882 | 0.2338411 |

Table 1: Convergence of real ($\sigma_r$) and imaginary part ($\sigma_i$) of eigen values for different $J$ and $M$. Other parameters: $Wo$ =15, $\delta$ =0.1, $f$ =0.1, $S = 5 \times 10^{-5}$, $k$=0.9 and $Re$=12000.

The numerical implementation is further validated by comparison with two limiting cases recovered from the present formulation. First, in the absence of temporal forcing, the system reduces to steady dusty gas channel flow, for which the stability characteristics agree with the results of Klinkenberg *et al.* (2011). Second, in the absence of particles, the formulation recovers the classical pulsatile channel flow, and the results are consistent with the asymptotic analysis of Tsigklifis & Lucey (2017). These benchmark comparisons, shown in figure 4, demonstrate excellent agreement and confirm the accuracy and robustness of the present numerical approach.

## 4. Results and discussion

We begin by examining the stability of steady particle-laden Poiseuille flow, which serves as a baseline for assessing the effect of pulsation. Previous studies have shown that particle-induced stability modifications depend non-monotonically on the relaxation time $S$, with very small $S$ producing weak destabilization and finite $S$ promoting stabilization through interphase momentum exchange (Klinkenberg *et al.* 2011; Rouquier *et al.* 2019). Building on this reference state, we investigate how time-periodic forcing modifies flow stability through changes in the neutral stability curves and disturbance growth. The pulsation amplitude is characterized by the flow-rate parameter $\delta$, with representative values $\delta = 0$, 0.1, and 0.2, where $\delta = 0$ corresponds to the steady-flow limit.

The influence of pulsation frequency is characterized by the Womersley number, which varied in the range of $6 \leq Wo \leq 25$, allowing us to assess the relative importance of oscillatory inertia and viscous diffusion. The analysis is performed for three representative particle relaxation times, $S = 10^{-7}$, $5 \times 10^{-5}$, and $2.5 \times 10^{-4}$, spanning regimes from a nearly instantaneous particle response to finite inertia effects. The particle loading is quantified through the mass fraction $f$, which is varied within the range $0 \leq f \leq 0.15$.

For each combination of parameters, we compute the neutral stability boundaries and the temporal growth rates of disturbances. Particular attention is given to the role of particle inertia and mass fraction, as well as the interaction between oscillatory forcing and interphase coupling. Using a systemically varying $\delta$, $Wo$, $S$, and $f$, the present analysis provides a comprehensive picture of how pulsation and particle dynamics jointly influence the onset and evolution of instability in a pulsatile channel flow.

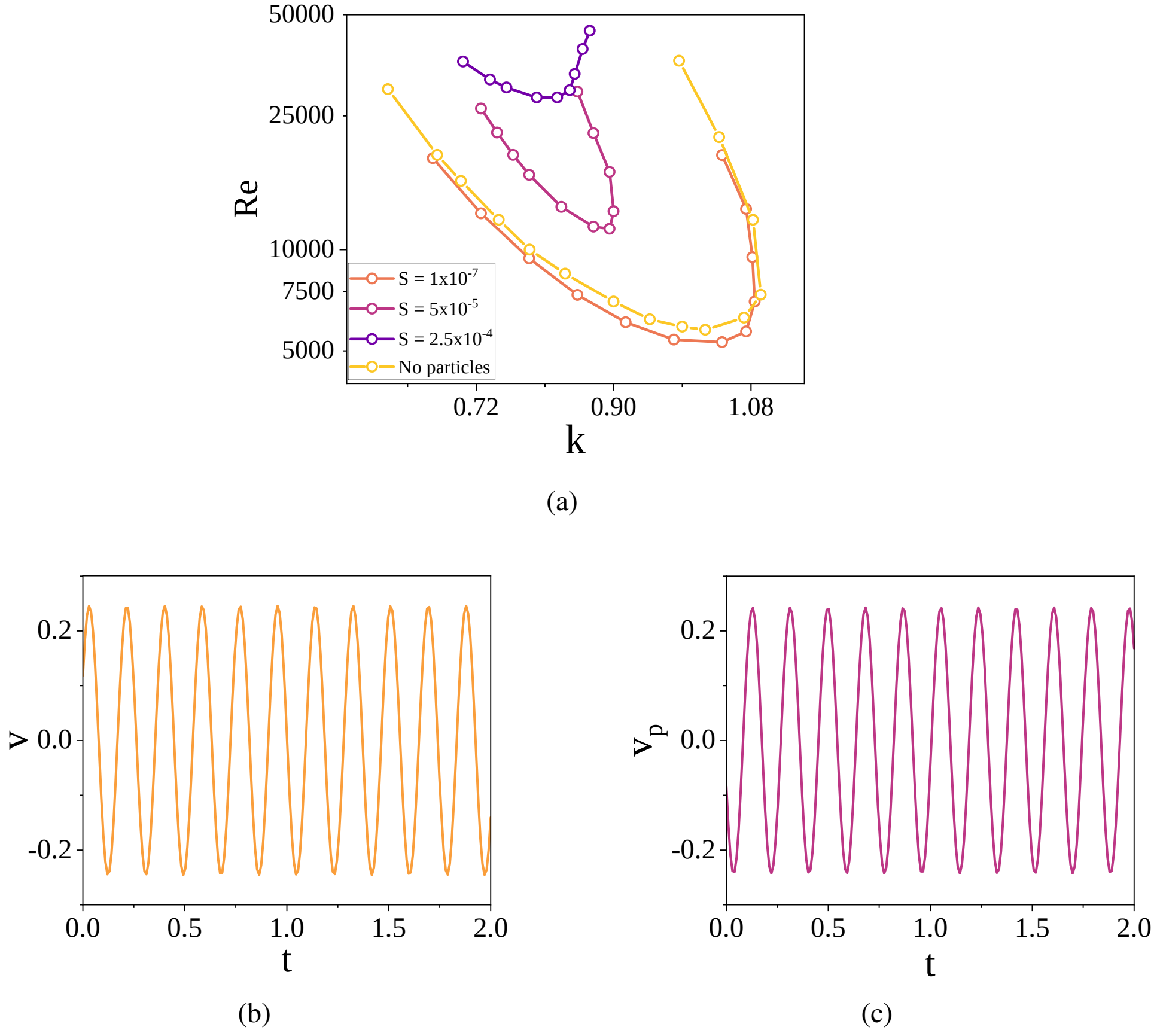

Figure 5: (a) Neutral stability curves for steady particle-laden Poiseuille flow at (f=0.1) for different particle relaxation times ($S$). (b,c) Temporal evolution of the wall normal perturbation velocities of the fluid, ($v$), and particle phase, ($v_p$), at ($y = 0.7$), evaluated at the critical conditions for ($S = 5 \times 10^{-5}$). In (b,c), the perturbation amplitudes are normalized by the initial perturbation kinetic energy. All quantities are dimensionless.

Unless stated otherwise, all quantities reported in this section are dimensionless. The velocity components are scaled by ($U_{\max}$), and the temporal traces of the perturbation amplitudes shown in figures 5 and 7 are normalized by the initial perturbation kinetic energy. The real part of the Floquet exponent, ($\sigma_r$), is used to assess temporal growth or decay.

### 4.1. *Steady Poiseuille flow with particles*

We first consider the stability of the Poiseuille flow laden with particles in the absence of pulsation ($\delta = 0$), which provides a baseline to understand the role of particle-fluid coupling prior to introducing oscillatory forcing. Figure 5a shows the neutral stability curves for a fixed particle mass fraction $f = 0.1$ and a varying particle relaxation time $S$. As shown in figure 5a, the results demonstrate that the influence of suspended particles on flow stability depends strongly on particle relaxation time. For a very short relaxation time ($S = 1 \times 10^{-7}$), the critical Reynolds number is slightly lower than that of the single-phase flow, indicating weak destabilization. In this regime, particles remain strongly coupled to the carrier phase and behave nearly as tracers. Since interphase slip remains negligible, particle addition primarily modifies the effective inertia of the suspension, which lowers the instability threshold. This behaviour is consistent with the classical stability analysis of dilute suspensions reported by Saffman (1962).

As the relaxation time increases ($S = 5 \times 10^{-5}$ and $S = 2.5 \times 10^{-4}$), the effect of suspended particles reverses and becomes stabilizing, as evidenced by the upward shift of the neutral curves and the increase in critical Reynolds number. In this finite-inertia regime, particles no longer respond instantaneously to fluid disturbances, and interphase slip develops between the particle and fluid phases. The resulting drag-mediated momentum exchange damps disturbance growth and shifts the onset of instability to larger Reynolds numbers. Physically, finite particle inertia weakens fluid–particle coupling and introduces an effective dissipative mechanism that suppresses disturbance amplification.

To further characterize the influence of particle inertia on disturbance dynamics, the temporal evolution of the wall-normal perturbation velocity is examined for both fluid and particle phases under critical conditions. The perturbations are normalized using their initial kinetic energy to ensure identical initial disturbance levels while preserving the relative temporal evolution of disturbance amplitudes. The critical Reynolds number, $Re_{cr}$ and the corresponding wavenumber $k_{cr}$ are obtained from the neutral stability curve for the representative case $S = 5 \times 10^{-5}$ and a mass fraction of $f = 0.1$. Figures 5b and 5c show the temporal evolution of the wall-normal perturbation velocities of the fluid and particle phases at $y = 0.7$, evaluated at the corresponding critical conditions. At criticality, both phases exhibit nearly neutrally stable oscillations, consistent with the marginal stability condition. The particle perturbation closely follows the fluid response, reflecting the strong interphase coupling characteristic of the present dilute suspension regime, although slight differences in amplitude evolution emerge due to finite particle inertia and drag coupling.

### 4.2. *Pulsatile Poiseuille flow with particles*

We now examine the stability of particle-laden channel flow subjected to a time-periodic pressure gradient, focusing on the combined effects of pulsation amplitude and frequency. The pulsation frequency is characterized by the Womersley number $Wo$, while the forcing amplitude is prescribed through the flow-rate parameter $\delta$. The particle mass fraction is fixed at $f = 0.1$, and representative relaxation times of $S = 10^{-7}, 5 \times 10^{-5}, 2.5 \times 10^{-4}$ are considered. Figure 6 summarizes the neutral stability curves for representative Womersley numbers and illustrates how pulsation modifies the onset of instability for varying particle inertia.

#### 4.2.1. *Low-frequency regime: Wo = 8*

We first consider a low-Womersley-number regime ($Wo$=8), corresponding to flow conditions where viscous diffusion can respond effectively over each oscillation cycle. Such conditions are also relevant to physiological pulsatile transport, including flow in the abdominal aorta (Lee *et al.* 2005; Asgharzadeh & Borazjani 2016).

As shown in figure 6a, increasing the pulsation amplitude $\delta$ destabilizes the flow across all particle relaxation times considered. The critical Reynolds number decreases and the unstable region broadens in wavenumber space, indicating that pulsation enhances both the onset and range of unstable disturbances. This behaviour is consistent with previous observations in single-phase pulsatile channel flow (Von Kerczek 1982; Singer *et al.* 1989; Straatman *et al.* 2002; Thomas *et al.* 2011). At low *Wo*, oscillatory forcing penetrates across most of the channel, producing temporally modulated shear throughout the bulk flow. Consequently, disturbances experience sustained interaction with the unsteady base flow over the oscillation cycle, enhancing disturbance amplification and lowering the instability threshold. Increasing $\delta$ intensifies this modulation, leading to the pronounced reduction in $Re_{cr}$ observed in figure 6a. Although a detailed disturbance-energy budget would be required to rigorously identify the balance between production and dissipation, the present trends strongly suggest that bulk oscillatory penetration promotes instability in this regime. The destabilizing influence of pulsation is strongest for the intermediate relaxation time $S = 5 \times 10^{-5}$, indicating enhanced sensitivity to oscillatory forcing when particle inertia becomes finite but fluid–particle coupling remains strong. Increasing $S$ nevertheless retains an overall stabilizing influence, consistent with the steady-flow results of §4.1, owing to enhanced interphase slip and drag-mediated damping.

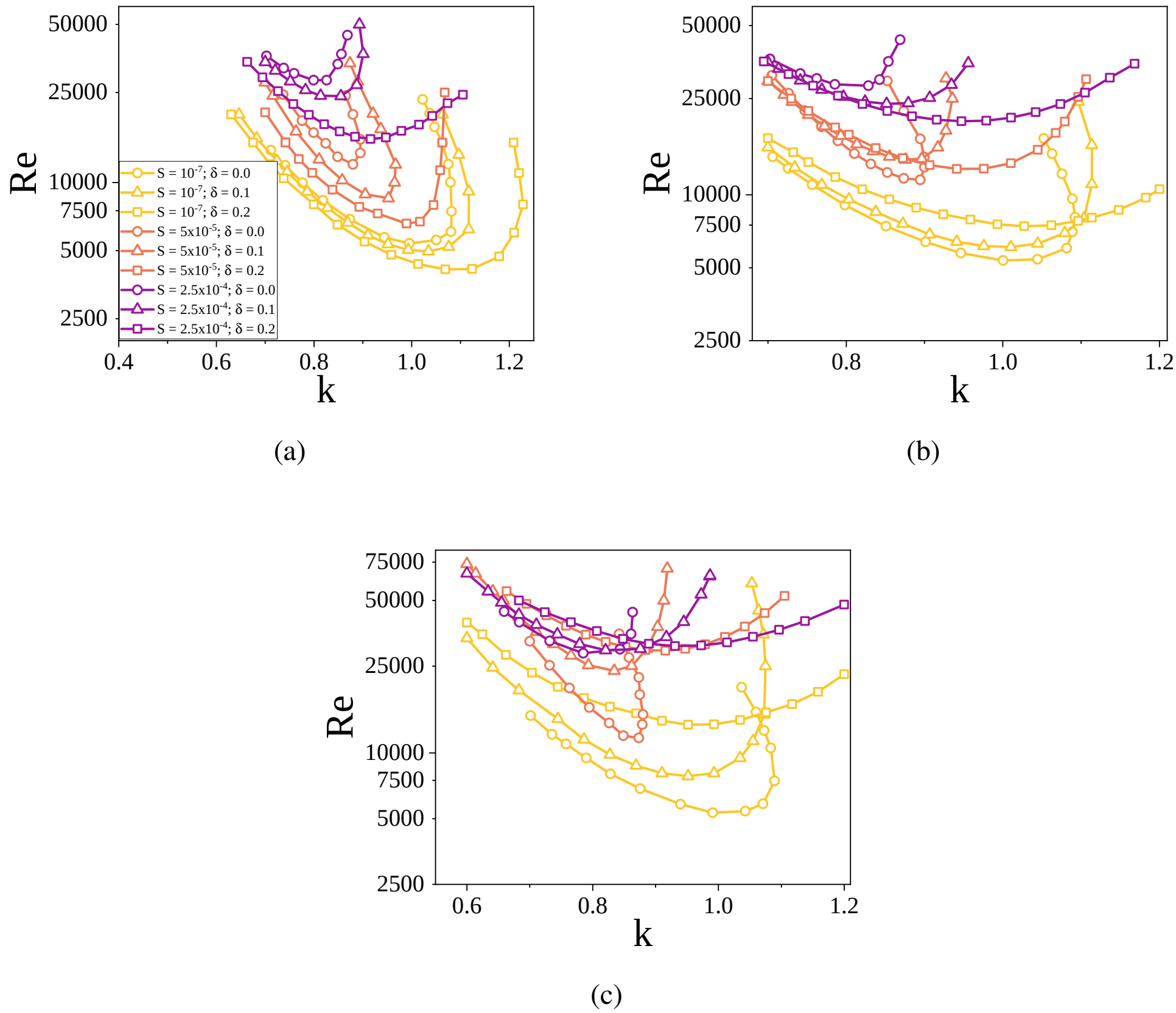

Figure 6: Neutral stability curves for particle-laden pulsatile Poiseuille flow for different particle relaxation times (S) and pulsation amplitudes ($\delta$): (a) (*Wo*=8), (b) (*Wo*=15), and (c) (*Wo*=20). Reynolds number and wavenumber are dimensionless. The curves have been computed from spectrally converged eigenvalue solutions.

Figures 7a and 7b show the temporal evolution of the normalized wall-normal perturbation velocities of the fluid and particle phases at $y = 0.7$. Since the oscillatory forcing penetrates deeply into the channel at $Wo = 8$, the interior location experiences substantial time-periodic modulation of the base flow. Both phases therefore exhibit relatively large-amplitude oscillations with strongly modulated temporal behaviour. The particle perturbation closely follows the fluid disturbance, reflecting the strong fluid–particle coupling characteristic of the present parameter range.

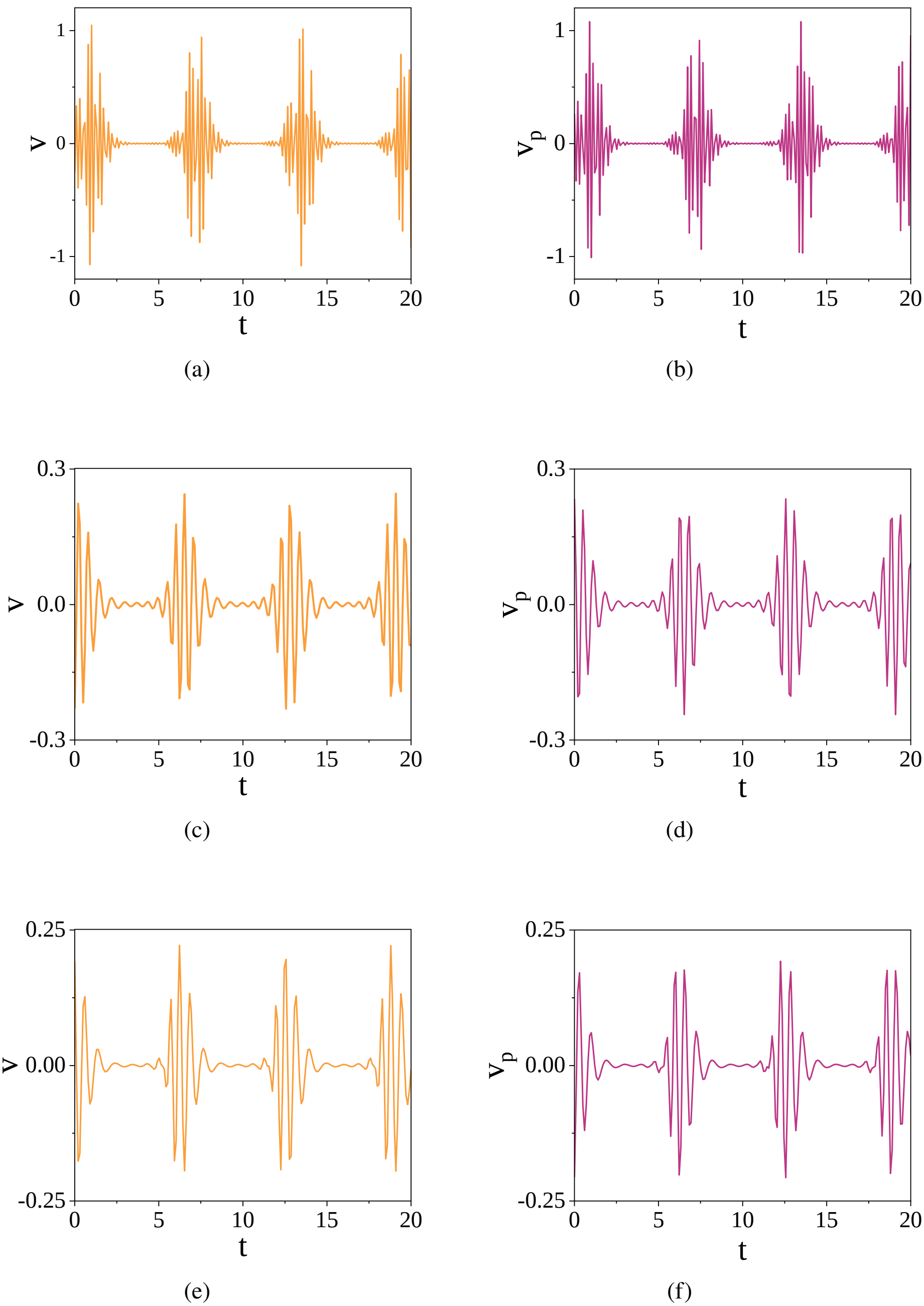

Figure 7: Temporal evolution of the wall normal perturbation velocities of the fluid, ($v$), and particle phase, ($v_p$), at ($y$ = 0.7): (a,b) ($Wo$=8), (c,d) ($Wo$=15), and (e,f) ($Wo$=20). In all panels, the perturbation amplitudes are normalized by the initial perturbation kinetic energy. Other parameters are ($f$=0.1), ($S = 5 \times 10^{-5}$), and ($\delta$=0.1). All quantities are dimensionless.

#### 4.2.2. *Intermediate-frequency regime: Wo=15*

At intermediate Womersley number ($Wo$=15), the influence of pulsation weakens substantially. As shown in figure 6b, the separation between neutral curves corresponding to different pulsation amplitudes decreases, indicating a reduction in the destabilizing influence observed at lower frequency. This behaviour reflects the reduced penetration of oscillatory motion into the channel interior. As $Wo$ increases, oscillatory forcing becomes progressively confined toward the near-wall region, weakening the temporal modulation experienced by disturbances in the channel core. Consequently, the effectiveness of the oscillatory base flow in sustaining disturbance amplification diminishes.

In this transitional regime, the net effect of pulsation becomes increasingly sensitive to particle dynamics. For small and intermediate relaxation times ($S = 1 \times 10^{-7}$ and $5 \times 10^{-5}$), increasing pulsation amplitude becomes weakly stabilizing. However, for larger relaxation time ($S = 2.5 \times 10^{-4}$), a slight destabilization persists, suggesting that finite particle inertia modifies the balance between oscillatory forcing and drag-mediated damping in a non-trivial manner.

The disturbance dynamics at $y = 0.7$, shown in figures 7c and 7d, exhibit smaller amplitudes than in the low-frequency case. Since the oscillatory penetration depth decreases with increasing $Wo$, disturbances in the channel interior experience weaker temporal modulation of the base flow. Nevertheless, both phases retain similar temporal structure, consistent with continued strong fluid–particle coupling in the present small-$SWo^2$ regime.

#### 4.2.3. *High-frequency regime: Wo=20*

At higher Womersley number ($Wo$=20), pulsation becomes clearly stabilizing. As shown in figure 6c, increasing pulsation amplitude raises the critical Reynolds number for all particle relaxation times considered, indicating systematic suppression of instability. At sufficiently higher $Wo$, unsteady inertial effects dominate viscous diffusion and confine the oscillatory motion to thin near-wall Stokes layers. The channel core therefore experiences only weak temporal modulation of the base flow, substantially reducing the ability of disturbances to extract energy from oscillatory forcing. As a result, disturbance amplification is suppressed and the instability threshold shifts to higher Reynolds number.

This behaviour is reflected in the disturbance evolution shown in figures 7e and 7f. At $y = 0.7$, which lies largely outside the region of strong oscillatory influence, the perturbation amplitudes are significantly reduced relative to the lower-$Wo$ cases. The particle phase continues to closely follow the fluid disturbance, indicating that fluid–particle coupling remains strong despite finite particle inertia. However, the localized nature of the oscillatory forcing substantially weakens the overall disturbance response. The results reveal a transition from pulsation-induced destabilization at low $Wo$ to stabilization at sufficiently high $Wo$. The transition is governed primarily by the penetration depth of oscillatory forcing: low-frequency pulsation modulates the shear throughout the bulk flow and enhances disturbance amplification, whereas high-frequency forcing becomes confined to near-wall regions and suppresses instability. Particle relaxation time and mass loading modify this transition through interphase momentum coupling and drag-mediated damping rather than through a resonance-like particle response.

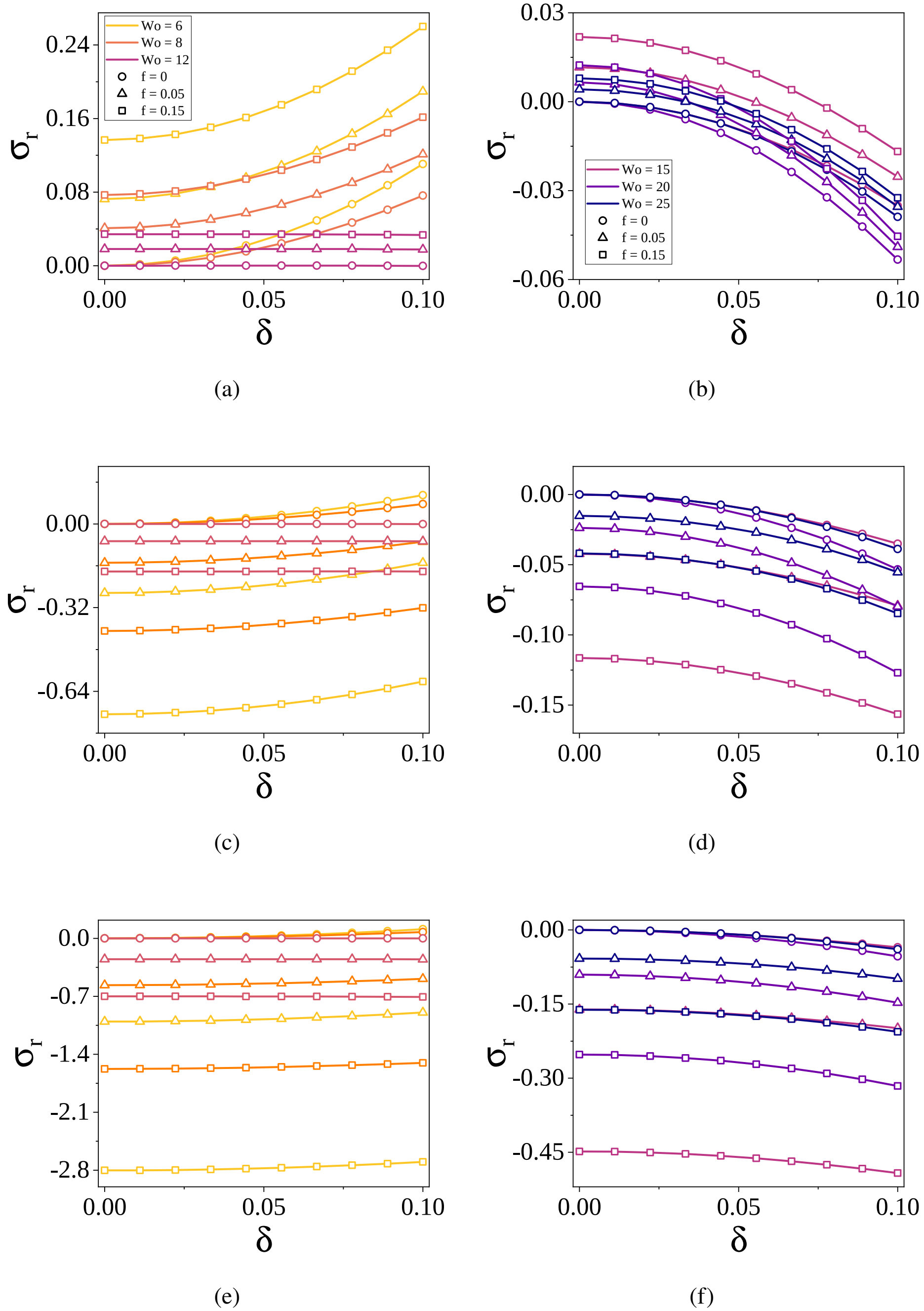

Figure 8: Real part of the Floquet growth rate, ($\sigma_r$), as a function of pulsation amplitude ($\delta$) for different particle mass fractions ($f$) and Womersley numbers ($Wo$), evaluated at fixed ($Re$=5772.22) and ($k$=1.0206). All quantities are dimensionless. (a) $S = 1 \times 10^{-7}$, $Wo$=6,8,12; (b) $S = 1 \times 10^{-7}$, $Wo$=15,20,25; (c) $S = 5 \times 10^{-5}$, $Wo$=6,8,12; (d) $S = 5 \times 10^{-5}$, $Wo$=15,20,25; (e) $S = 2.5 \times 10^{-4}$, $Wo$=6,8,12; (f) $S = 2.5 \times 10^{-4}$, $Wo$=15,20,25.

### 4.3. *Effect of particle mass fraction*

We now examine the influence of the particle mass fraction, $f$, on the stability of the pulsatile channel flow. In steady particle-laden channel flow, suspended particles may either stabilize or destabilize the flow depending on particle relaxation time $S$ (Klinkenberg *et al.* 2011). As shown above, particles with very short relaxation times ($S = 1 \times 10^{-7}$) remain strongly coupled to the carrier phase and promote destabilization through increased effective suspension inertia, whereas particles with larger relaxation times ($S = 5{\times}10^{-5}$ and $2.5{\times}10^{-4}$) exert a stabilizing influence through enhanced interphase slip and drag-mediated damping.

In the pulsatile system, the influence of particle loading is further modified by oscillatory forcing, characterized by the Womersley number $Wo$. To isolate the effect of particle loading, we examined the temporal growth rate $\sigma_r$ at fixed $Re = 5772.22$ and $k = 1.0206$, corresponding to the critical conditions of single-phase steady Poiseuille flow. The pulsation amplitude is varied over $0 \leq \delta \leq 0.1$, while $6 \leq Wo \leq 25$. Variations in $\sigma_r$ therefore reflect local changes in disturbance amplification at fixed ($Re$, $k$) and should be distinguished from the global stability trends based on the critical Reynolds number discussed in §4.4.

#### 4.3.1. *Small relaxation time: $S = 1 \times 10^{-7}$*

For very small relaxation time ($S = 1 \times 10^{-7}$), particles remain strongly coupled to the carrier flow and behave approximately as tracers. Figures 8a and 8b show that increasing particle loading increases the disturbance growth rate across all Womersley numbers considered, indicating a destabilizing influence of suspended particles. This behaviour is consistent with the steady-flow limit and reflects the increase in effective suspension inertia associated with stronger particle loading (Saffman 1962).

At low Womersley number ($Wo = 6, 8$), the growth rate increases strongly with pulsation amplitude, indicating pulsation-induced destabilization for all mass fractions. In this regime, oscillatory forcing penetrates across most of the channel, producing temporally modulated shear throughout the bulk flow and enhancing disturbance amplification. Increasing particle loading amplifies this destabilizing effect by strengthening the inertial response of the suspension while maintaining strong fluid–particle coupling. As $Wo$ increases toward intermediate values ($Wo \approx 12$), the sensitivity of $\sigma_r$ to pulsation amplitude weakens, indicating a transition between destabilizing and stabilizing regimes. At sufficiently high Womersley number ($Wo \geq 15$), increasing $\delta$, treduces the growth rate and eventually suppresses instability entirely ($\sigma_r < 0$). In this regime, oscillatory motion becomes increasingly confined to near-wall Stokes layers, substantially reducing the penetration of unsteady forcing into the channel interior. Consequently, disturbance amplification weakens and the relative influence of particle loading diminishes.

#### 4.3.2. *Intermediate relaxation time: $S = 5 \times 10^{-5}$*

For intermediate relaxation time ($S = 5 \times 10^{-5}$), particle loading produces a stabilizing influence across all Womersley numbers, as shown in figures 8c and 8d. Unlike the tracer-like regime, finite particle inertia allows interphase slip to develop, increasing drag-mediated momentum exchange and suppressing disturbance growth. At low Womersley numbers ($Wo$= 6,8), pulsation remains destabilizing, as reflected by the increase in growth rate with increasing pulsation amplitude. However, the destabilizing influence of oscillatory forcing is now partially counteracted by particle-induced damping. In this regime, oscillatory penetration extends across much of the channel and disturbances continue to experience strong temporal modulation of the base flow.

At larger Womersley number ($Wo \geq 15$), increasing pulsation amplitude reduces the disturbance growth rate, indicating stabilization. Since oscillatory forcing becomes increasingly confined to near-wall layers, disturbances in the channel core experience weaker temporal modulation. Combined with drag-mediated damping arising from finite particle inertia, this confinement promotes suppression of instability. The influence of particle mass fraction is strongest at intermediate Womersley numbers, where the competing effects of oscillatory penetration and drag-mediated damping are most balanced.

#### 4.3.3. *Large relaxation time:* $S = 2.5 \times 10^{-4}$

For larger relaxation time ($S = 2.5 \times 10^{-4}$), the qualitative behaviour remains similar to the intermediate-$S$ case but with substantially greater sensitivity to particle loading. Increasing $f$ consistently reduces the disturbance growth rate across all Womersley numbers considered, indicating stronger stabilization by suspended particles. This enhanced sensitivity reflects the greater importance of finite particle inertia, which increases interphase slip and strengthens drag-mediated damping of disturbances. Although pulsation continues to destabilize the flow at low *Wo* and stabilize it at higher *Wo*, the stabilizing influence of particle loading becomes substantially more pronounced than in the intermediate-relaxation-time regime. Therefore, the effect of particle mass fraction depends strongly on both particle relaxation time and pulsation frequency. At small $S$, increasing particle loading enhances effective suspension inertia and promotes destabilization. At larger $S$, increasing particle loading strengthens drag-mediated damping and stabilizes the flow. The influence of $f$ is most pronounced at intermediate Womersley numbers, where oscillatory penetration and particle-induced damping are most strongly balanced.

### 4.4. *Variation of critical Reynolds number*

The critical Reynolds number, $Re_{cr}$, defined as the Reynolds number at which the maximum temporal growth rate vanishes, serves as a global measure of flow stability. Its dependence on pulsation amplitude $\delta$, Womersley number *Wo*, particle relaxation time $S$, and mass fraction $f$ is examined in figure 9.

In the steady limit ($\delta = 0$), the present results are consistent with established trends for particle-laden Poiseuille flow. For small relaxation time $S = 10^{-7}$, particles remain strongly coupled to the carrier phase and behave nearly as tracers. In this regime, $Re_{cr}$ decreases relative to the single-phase value, indicating destabilization associated with increased effective suspension inertia (Klinkenberg *et al.* 2011). By contrast, for larger relaxation time, $S = 2.5 \times 10^{-4}$, finite particle inertia promotes interphase slip and drag-mediated damping, increasing $Re_{cr}$ and stabilizing the flow.

Pulsatile forcing modifies these steady-flow trends a strongly frequency-dependent manner. For $S = 1 \times 10^{-7}$ and $f = 0.05$, figure 9a shows that $Re_{cr}$ decreases with increasing pulsation amplitude at low Womersley numbers (*Wo*=6–8), indicating pulsation-induced destabilization. In this regime, oscillatory forcing penetrates across most of the channel, producing temporally modulated shear throughout the bulk flow and enhancing disturbance amplification. Within the range considered, the strongest destabilization occurs near *Wo*=8. At intermediate *Wo*=(10-12), the sensitivity of $Re_{cr}$ to pulsation amplitude weakens, marking a transition between destabilizing and stabilizing regimes. For $Wo \geq 15$, the trend reverses and $Re_{cr}$ increases with $\delta$, indicating stabilization. This reversal reflects the confinement of oscillatory motion to thin near-wall Stokes layers at larger *Wo*, which reduces the penetration of time-periodic shear into the channel core and suppresses disturbance amplification. The strongest stabilization occurs at *Wo*=25, the highest Womersley number considered.

Increasing the particle relaxation time to $S = 2.5 \times 10^{-4}$ preserves the qualitative transition from low-Wo destabilization to high-Wo stabilization, but shifts the location and magnitude of the response. As shown in figure 9b, the strongest destabilization shifts from *Wo*=8 for $S = 10^{-7}$ to approximately *Wo*=10 for the larger relaxation time. This shift reflects the increased importance of interphase momentum exchange at finite particle inertia, which modifies the balance between oscillatory penetration and drag-mediated damping. For *Wo* $\geq$ 15, pulsation again stabilizes the flow, with the largest increase in $Re_{cr}$ occurring at *Wo*=25.

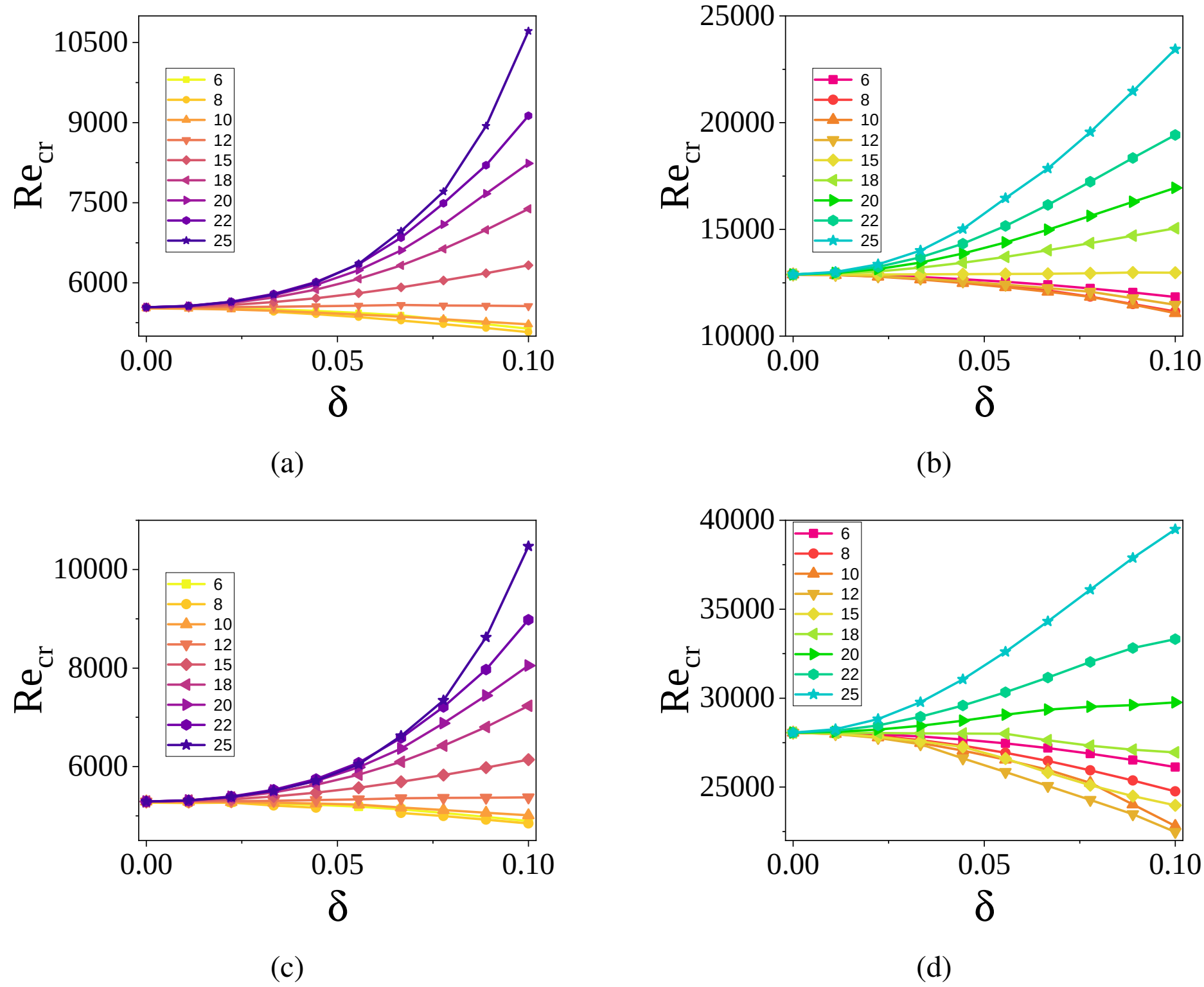

Figure 9: Variation of the critical Reynolds number, $Re_{cr}$, with pulsation amplitude $\delta$ for different Womersley numbers and particle parameters. (a) $f = 0.05$, $S = 1 \times 10^{-7}$; (b) $f = 0.05$, $S = 2.5 \times 10^{-4}$; (c) $f = 0.1$, $S = 1 \times 10^{-7}$; (d) $f = 0.1$, $S = 2.5 \times 10^{-4}$. The results show the transition from pulsation-induced destabilization at low *Wo* to stabilization at high *Wo*, and the modification of this transition by particle inertia and mass loading.

Figures 10a and 10b further isolate the role of particle mass fraction by showing the variation of $Re_{cr}$ with pulsation amplitude at selected Womersley numbers. For $S = 10^{-7}$ , increasing $f$ lowers $Re_{cr}$ across all Womersley numbers, demonstrating that the destabilizing influence of tracer-like particles persists in both destabilizing and stabilizing pulsation regimes. Thus, although high-*Wo* pulsation suppresses instability, particle loading continues to shift the critical threshold downward through increased effective suspension inertia. For $S = 2.5{\times}10^{-4}$, increasing $f$ raises $Re_{cr}$ at all Womersley numbers considered, confirming that finite-inertia particles stabilize the flow through drag-mediated attenuation of disturbances. In this regime, the stabilizing influence of particle loading reinforces the stabilizing effect of high-frequency pulsation.

To provide a unified description of the combined effects of particle inertia and pulsation frequency, the net effect of pulsation is quantified using

$$\Delta Re_{cr} = Re_{cr}(\delta = 0.1) - Re_{cr}(\delta = 0), \tag{4.1}$$

where positive and negative values correspond to stabilization and destabilization, respec-tively. Figures 11a and 11b present phase diagrams of $\Delta Re_{cr}$ in the ($Wo$,$S$) parameter space for $f = 0.05$ and $f = 0.1$. At the particle mass fractions increases from $f = 0.05$ to $f = 0.1$, the transition boundary systematically shifts toward larger Womersley number for all values of $S$. Consequently, the destabilizing region expands and higher forcing frequencies are required to achieve stabilization. This shift does not reflect a resonance-like particle response; rather, it arises from the manner in which particle loading modifies the stability threshold through effective suspension inertia at small $S$ and drag-mediated damping at larger $S$. In general, the critical Reynolds number results demonstrate that the transition from pulsation-induced destabilization to stabilization is governed primarily by the Womersley-number dependence of oscillatory penetration. Particle relaxation time and mass fraction do not alter the underlying mechanism, but systematically shift the transition through interphase momentum exchange. At small $S$, particles remain strongly coupled to the carrier flow and promote destabilization through increased effective inertia, whereas at larger $S$, finite interphase slip enhances drag-mediated damping and stabilizes the flow.

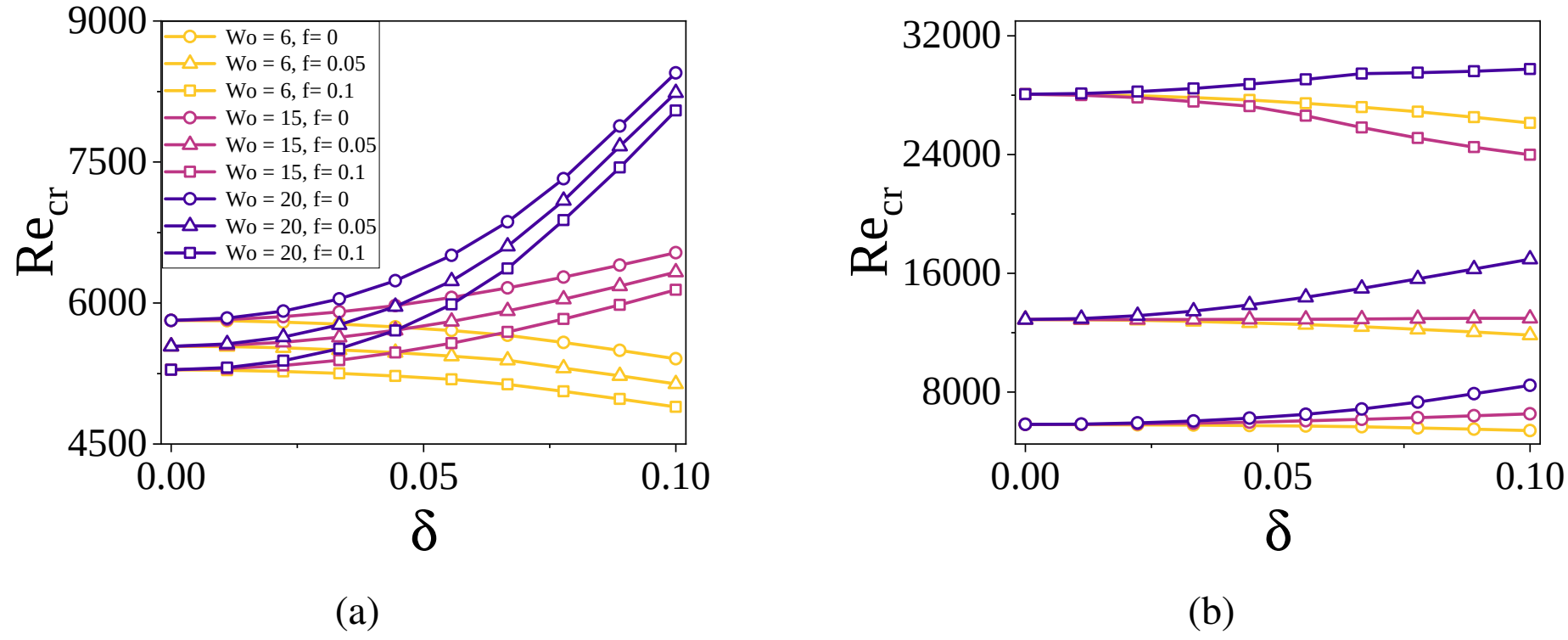

Figure 10: Variation of critical Reynolds number, $Re_{cr}$ with pulsation amplitude $\delta$ for different particle mass fraction $f$. (a) $S = 1 \times 10^{-7}$; (b) $S = 2.5 \times 10^{-4}$.

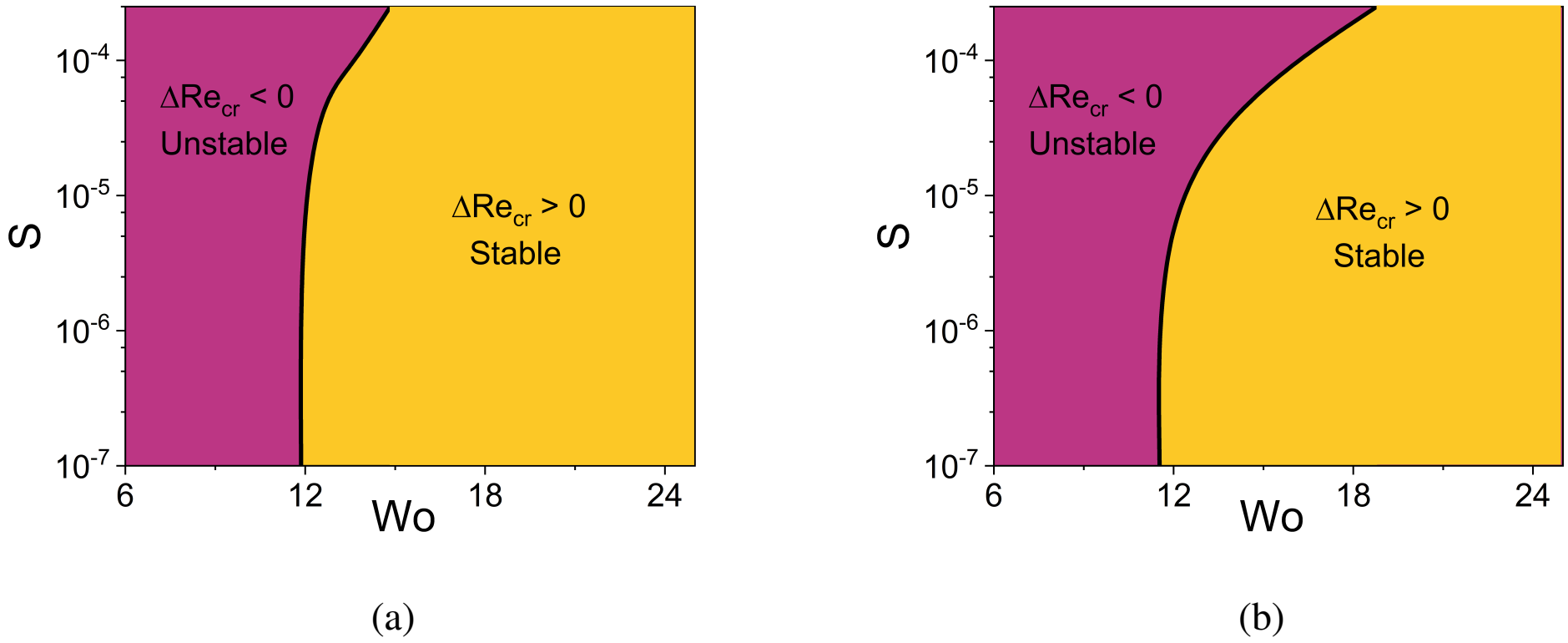

Figure 11: Phase diagrams showing the variation of $\Delta Re_{cr}$ with $Wo$ (6 -25) and $S$ ($10^{-7}$ to $2.5 \times 10^{-4}$) for $\delta = 0.1$ and different $f$: (a) $f = 0.05$, and (b) $f = 0.1$.

## 5. Conclusions

The linear stability of pulsatile particle-laden channel flow has been investigated using a Floquet-based framework. The suspension was modeled as a dilute two-phase system in which interphase momentum transfer is governed by Stokes drag, while the base flow is driven by a time-periodic pressure gradient. By formulating the stability problem in terms of time-periodic eigenmodes, the present approach captures the coupled influence of oscillatory forcing, particle relaxation time, and particle loading on disturbance amplification and instability onset.

In the steady-flow limit, the results recover the established behaviour of particle-laden Poiseuille flow and provide a consistent baseline for interpreting the effects of pulsatile forcing. Particles with very small relaxation time ($S = 10^{-7}$) remain strongly coupled to the carrier flow and behave nearly as tracers, producing weak destabilization through increased effective suspension inertia. In contrast, particles with finite relaxation time ($S = 5\times10^{-5}$ and $2.5\times10^{-4}$) stabilize the flow through interphase slip and drag-mediated damping, increasing the critical Reynolds number relative to the single-phase limit. These results confirm that the sign and magnitude of the particle contribution depend fundamentally on the strength of fluid–particle coupling.

The introduction of pulsatile forcing produces a qualitatively distinct and strongly frequency-dependent stability response. At low Womersley number ($Wo \approx 6$–$8$), increasing pulsation amplitude destabilizes the flow, reducing the critical Reynolds number and broadening the range of unstable disturbances. This behaviour arises because oscillatory forcing penetrates across much of the channel, generating temporally modulated shear throughout the bulk flow and enhancing disturbance amplification. At sufficiently large Womersley number ($Wo \approx 20$–$25$), the trend reverses and pulsation stabilizes the flow. In this regime, oscillatory motion becomes confined to thin near-wall Stokes layers, reducing temporal modulation within the channel interior and suppressing disturbance growth. The transition between these two regimes occurs at intermediate Womersley number ($Wo \approx 12$–$15$), marking a shift from bulk-dominated oscillatory interaction to near-wall-confined forcing.

Particle relaxation time and particle loading systematically modify this transition. For tracer-like particles ($S = 10^{-7}$), increasing mass fraction lowers the critical Reynolds number and amplifies destabilization through increased effective suspension inertia. In contrast, for particles with finite inertia ($S = 5\times10^{-5}$ and $2.5\times10^{-4}$), increasing particle loading enhances drag-mediated damping and promotes stabilization. Although the underlying transition between destabilization and stabilization remains controlled primarily by Womersley number, particle inertia and loading shift the stability boundary through interphase momentum exchange.

The phase diagram analysis based on $\Delta Re_{cr} = Re_{cr}(\delta = 0.1) - Re_{cr}(\delta = 0)$, provides a unified interpretation of these effects. Two distinct regimes emerge in the ($Wo$,$S$) parameter space: a low-$Wo$ destabilizing regime associated with strong oscillatory penetration and a high-$Wo$ stabilizing regime associated with near-wall confinement of oscillatory motion. Increasing particle loading systematically shifts the transition boundary toward larger Womersley number, expanding the destabilizing regime and requiring stronger oscillatory forcing to achieve stabilization. Importantly, the corresponding values of $SWo^2$ along the transition boundary remain small throughout the parameter range considered, indicating that particles remain strongly coupled to the carrier flow and that the observed transition is not governed by a resonance-like particle response. The principal finding of this work is that the transition between pulsation induced destabilization and stabilization is governed by the coupled effects of oscillatory penetration and particle-fluid momentum exchange. While the Womersley number dependence of oscillatory penetration determines whether pulsation destabilizes or stabilizes the flow, particle relaxation time and mass loading systematically shift the instability threshold through effective suspension inertia and drag mediated damping. Consequently, instability in pulsatile suspensions cannot be understood from oscillatory forcing or particle dynamics in isolation, but instead emerges from their coupled interaction.

These findings have direct implications for pulsatile multiphase systems, including transport of suspended cells and particulates in vascular environments, drug and aerosol delivery under oscillatory flow conditions, and engineering systems involving periodically forced suspensions. Beyond the present theoretical predictions, the results motivate targeted experiments to directly characterize instability onset in pulsatile particle-laden flows. Time-resolved measurements using particle image velocimetry (PIV) and particle tracking velocimetry (PTV) in controlled oscillatory channel configurations could quantify disturbance growth, particle-fluid slip, and the transition between destabilizing and stabilizing regimes across varying pulsation frequencies and particle properties. More broadly, the Floquet-based framework developed here provides a rigorous foundation for future extensions to concentrated suspensions with stronger interphase coupling, nonlinear disturbance dynamics, migration effects, spatially non-uniform particle distributions, and non-Newtonian carrier fluids relevant to biological and industrial systems. By linking oscillatory forcing with particle dynamics, the present work establishes a predictive framework for understanding instability mechanisms in realistic pulsatile multiphase flows.

**Funding.** P.M. and A.R. were supported by the National Science Foundation-CBET (Award no. 2230892) and (Award no. 2335195).

**Declaration of interests.** The authors report no conflict of interest.

## Appendix A. Applicability of Squire's theorem for particle-laden pulsatile flow

To assess the dimensionality of the least stable disturbances, we consider three-dimensional perturbations of the form $\xi'(x, y, z, t) = \hat{\xi}(y, t)e^{i(\alpha x+\beta z)}$, where $\hat{\xi}(y) = (\hat{u}, \hat{v}, \hat{w}, \hat{u}_p, \hat{v}_p, \hat{w}_p, \hat{n}, \hat{p})$ denotes the amplitudes of the fluid velocity, particle velocity, particle number density, and pressure perturbations.

Substitution into the linearized governing equations yields the following system. The fluid-phase continuity equation is

$$i\alpha\hat{u} + D\hat{v} + i\beta\hat{w} = 0 \tag{A 1}$$

The fluid momentum equations become

$$\frac{Wo^2}{Re}\frac{\partial\hat{u}}{\partial t}+Ui\alpha\hat{u}+U'\hat{v} = -i\alpha\hat{p}+\frac{1}{Re}\left(D^2-(\alpha^2+\beta^2)\right)\hat{u}-\frac{fN_0}{SRe}(\hat{u}-\hat{u}_p)-\frac{f}{SRe}(U-U_p)\hat{n} \tag{A 2}$$

$$\frac{Wo^2}{Re}\frac{\partial\hat{v}}{\partial t} + Ui\alpha\hat{v} = -\frac{\partial\hat{p}}{\partial y} + \frac{1}{Re}\left(D^2-(\alpha^2+\beta^2)\right)\hat{v} - \frac{fN_0}{SRe}(\hat{v}-\hat{v}_p) \tag{A 3}$$

$$\frac{Wo^2}{Re}\frac{\partial\hat{w}}{\partial t} + Ui\alpha\hat{w} = -i\beta\hat{p} + \frac{1}{Re}\left(D^2-(\alpha^2+\beta^2)\right)\hat{w} - \frac{fN_0}{SRe}(\hat{w}-\hat{w}_p) \tag{A 4}$$

The particle phase equations can be reformulated as

$$\frac{Wo^2}{Re}\frac{\partial\hat{n}}{\partial t} + U_p i\alpha\hat{n} + i\alpha\hat{u}_p + \frac{\partial\hat{v}_p}{\partial y} + i\beta\hat{w}_p = 0, \tag{A 5}$$

$$\frac{Wo^2}{Re}\frac{\partial \hat{u}_p}{\partial t} + U_p i\alpha\hat{u}_p + U_p'\hat{v}_p = \frac{1}{SRe}(\hat{u} - \hat{u}_p) \tag{A 6}$$

$$\frac{Wo^2}{Re}\frac{\partial \hat{v}_p}{\partial t} + U_p i\alpha\hat{v}_p = \frac{1}{SRe}(\hat{v} - \hat{v}_p) \tag{A 7}$$

$$\frac{Wo^2}{Re}\frac{\partial \hat{w}_p}{\partial t} + U_p i\alpha\hat{w}_p = \frac{1}{SRe}(\hat{w} - \hat{w}_p) \tag{A 8}$$

In the above equations, different parameters are defined as $D = \frac{\partial}{\partial y}$, $D^2 = \frac{\partial^2}{\partial y^2}$, $U' = dU/dy$ and $U_p' = dU_p/dy$ . No-slip and no-penetration boundary conditions are imposed for both phases:

$$\begin{aligned} \hat{u} = \hat{v} = \hat{w} = 0 \quad \text{at } y = \pm 1 \\ \hat{u}_p = \hat{v}_p = \hat{w}_p = 0 \quad \text{at } y = \pm 1 \end{aligned} \tag{A 9}$$

In the present formulation, the base-state velocities are assumed identical, $U = U_p$. Under this assumption, the interphase coupling term proportional to $(U - U_p)\hat{n}$ vanishes. Consequently, the particle number density perturbation $\hat{n}$ does not appear in the momentum equations and evolves independently through the particle continuity equation. At the linear level, $\hat{n}$ therefore behaves as a passive scalar and does not influence the stability of the velocity field. It can thus be omitted from the subsequent analysis.

To examine the dimensionality of the least stable modes, a Squire-type transformation is introduced. Defining $\tilde{k} = \sqrt{\alpha^2 + \beta^2}$, $\alpha Re = \tilde{k}\tilde{Re}$, and introducing transformed variables $\alpha\hat{u} + \beta\hat{w} = \tilde{k}\tilde{u}$, $\hat{v} = \tilde{v}$, $\tilde{p} = k\hat{p}/\alpha$, $\alpha\hat{u}_p + \beta\hat{w}_p = \tilde{k}\tilde{u}_p$, $\hat{v}_p = \tilde{v}_p$, the three-dimensional disturbance problem can be mapped onto an equivalent two-dimensional system.

Under this transformation, the governing equations retain the same structure as the corresponding two-dimensional stability equations, with streamwise wavenumber $\tilde{k}$ and Reynolds number $\tilde{Re}$. The transformed problem is therefore mathematically equivalent to a two-dimensional formulation, given by

$$\mathrm{i}\tilde{k}\tilde{u} + D\tilde{v} = 0, \tag{A 10}$$

$$\frac{Wo^2}{\tilde{Re}}\frac{\partial \tilde{u}}{\partial t} + Uik\tilde{u} + \frac{dU}{dy}\tilde{v} = -ik\tilde{p} + \frac{1}{\tilde{Re}}\left(D^2 - \tilde{k}^2\right)\tilde{u} - \frac{fN_0}{S\tilde{Re}}(\tilde{u} - \tilde{u}_p) \tag{A 11}$$

$$\frac{Wo^2}{\tilde{Re}}\frac{\partial \tilde{v}}{\partial t} + Uik\tilde{v} = -\frac{\partial \tilde{p}}{\partial y} + \frac{1}{\tilde{Re}}\left(D^2 - \tilde{k}^2\right)\tilde{v} - \frac{fN_0}{S\tilde{Re}}(\tilde{v} - \tilde{v}_p) \tag{A 12}$$

$$\frac{Wo^2}{\tilde{Re}}\frac{\partial \tilde{u}_p}{\partial t} + U_p ik\tilde{u}_p + \frac{\partial U_p}{\partial y}\tilde{v}_p = \frac{1}{SRe}(\tilde{u} - \tilde{u}_p) \tag{A 13}$$

$$\frac{Wo^2}{\tilde{Re}}\frac{\partial \tilde{v}_p}{\partial t} + U_p ik\tilde{v}_p = \frac{1}{S\tilde{Re}}(\tilde{v} - \tilde{v}_p) \tag{A 14}$$

$$\begin{aligned} \tilde{u} = \tilde{v} = 0 \quad \text{at } y = \pm 1 \\ \tilde{u}_p = \tilde{v}_p = 0 \quad \text{at } y = \pm 1 \end{aligned} \tag{A 15}$$

For a fixed streamwise wavenumber $\alpha$, the transformed Reynolds number satisfies $\tilde{Re} \leq$ Re. This implies that the smallest critical Reynolds number occurs when the spanwise wavenumber vanishes ($\beta$=0). Therefore, the most unstable disturbances in particle-laden pulsatile channel flow are two-dimensional, and the stability analysis may be restricted to two-dimensional perturbations without loss of generality. This result is consistent with classical Squire-type arguments for parallel shear flows and extends their applicability to particle-laden pulsatile systems under the present modelling assumptions. A rigorous extension of Squire's theorem to time-periodic (Floquet) systems is non-trivial; however, the present transformation demonstrates that the least stable modes remain two-dimensional under the assumptions adopted here.